\documentclass{JFM-FLM_Au}

\lefttitle{Li-Sheng Jiang, Ao Xu and Heng-Dong Xi}
\righttitle{Journal of Fluid Mechanics}

\title{Wall-modelled large-eddy simulation of turbulent channel flow with unstable stratification}

\author{Li-Sheng Jiang\aff{1}, Ao Xu\aff{1,2} \and Heng-Dong Xi\aff{1,2}}
\affiliation{\aff{1}Institute of Extreme Mechanics, School of Aeronautics, Northwestern Polytechnical University, Xi'an 710072, PR China
\aff{2} National Key Laboratory of Aircraft Configuration Design, Key Laboratory for Extreme Mechanics of Aircraft of Ministry of Industry and Information Technology, Xi'an 710072, PR China}

\corresau{Ao Xu, \email{axu@nwpu.edu.cn}}

\makeatletter
\newcommand{\JFMresumeafterdisplay}{%
  \@ifnextchar\par{}{\noindent}%
}
\makeatother
\AtBeginDocument{%
  \BeforeBeginEnvironment{equation}{\par}%
  \AfterEndEnvironment{equation}{\JFMresumeafterdisplay}%
  \BeforeBeginEnvironment{equation*}{\par}%
  \AfterEndEnvironment{equation*}{\JFMresumeafterdisplay}%
  \BeforeBeginEnvironment{align}{\par}%
  \AfterEndEnvironment{align}{\JFMresumeafterdisplay}%
}

\begin{document}
\maketitle

\begin{abstract}
Unstable thermal stratification modifies near-wall momentum and heat transport, causing the mean velocity profile to depart from the classical logarithmic law and complicating wall modelling for turbulent mixed convection.
We develop a buoyancy-modified logarithmic-quadratic wall model for incompressible Poiseuille--Rayleigh--B\'enard (PRB) flow.
The model combines an approximately linear relation between the near-wall mean temperature and mean streamwise velocity with a thermally modified mean-gradient representation inspired by mixing-length scaling.
\emph{A priori} assessments using wall quantities from the direct numerical simulation (DNS) database of \citet{PirozzoliBernardiniVerziccoOrlandi2017} show that the calibrated wall law reconstructs near-wall velocity and wall-function-equivalent eddy-viscosity profiles.
We implement the wall model in wall-modelled large-eddy simulations (WMLES) at friction Reynolds numbers up to $Re_\tau\approx6000$ and Rayleigh numbers up to $Ra=10^{10}$.
For cases with DNS reference profiles at $Ra=10^8$ and $10^9$, the maximum pointwise absolute relative errors are $3.6\%$ for the mean velocity and $1.9\%$ for the mean temperature.
For cases with available DNS global-transport data, the maximum relative deviations in the Nusselt number $Nu$ and skin-friction coefficient $C_f$ are $11.7\%$ and $15.9\%$, respectively.
The WMLES reduces the mesh count by factors of approximately $195$--$542$ relative to the corresponding DNS meshes.
For the $Ra=10^{10}$ and $Ri_b=0.1$ case, extrapolation of reference DNS resolution strategies gives a mesh count of order $10^{11}$, approximately three orders of magnitude larger than the present WMLES mesh count.
We also examine how the balance between shear and buoyancy reorganises flow structure, and we identify signatures consistent with the coexistence of streamwise-elongated motions resembling very-large-scale motions (VLSMs) and buoyancy-associated streamwise rolls.
\end{abstract}



\section{Introduction}\label{sec:introduction}

Mixed convection, driven by the interplay between shear and buoyancy, is ubiquitous in both natural and engineering flows.
In nuclear-reactor cooling channels, pressure-driven coolant flow may interact with buoyancy induced by wall heating, modifying turbulence and heat transfer \citep{CottonIsmaelKirwin2001}.
In the atmospheric boundary layer, mean wind shear interacts with buoyancy generated by surface heating, producing flow structures that range from shear-dominated streaks to streamwise convective rolls \citep{SaleskyAnderson2018,JayaramanBrasseur2021}.
In such flows, buoyancy modifies the velocity field, while the resulting fluid motions redistribute heat.
This coupling can alter both heat transfer and the spatial organisation of turbulence as the relative strengths of shear and buoyancy vary.

A canonical configuration for studying mixed convection is the Poiseuille--Rayleigh--B\'enard (PRB) system \citep{PirozzoliBernardiniVerziccoOrlandi2017,MadhusudananIllingworthMarusicChung2022,XuLiXi2025}, which combines pressure-driven Poiseuille flow with buoyancy-driven Rayleigh--B\'enard (RB) convection.
At fixed Prandtl number $Pr$, the PRB system is governed by the bulk Reynolds number $Re_b$ and the Rayleigh number $Ra$, while the resulting wall stress is characterised by the friction Reynolds number $Re_\tau$.
The DNS database of \citet{PirozzoliBernardiniVerziccoOrlandi2017} extends to $Re_\tau=946.41$ and $Ra=10^9$, and, to the best of our knowledge, these remain the highest Reynolds and Rayleigh numbers attained in DNS of the canonical PRB configuration.
Access to still higher Reynolds and Rayleigh numbers would provide greater scale separation and enable the investigation of multiscale interactions relevant to geophysical and industrial flows, but DNS under such conditions remains computationally prohibitive.

Large-eddy simulation (LES) balances accuracy and computational cost by resolving the large energy-containing eddies while modelling the smaller subgrid-scale (SGS) motions.
However, wall-resolved LES must still resolve the steep gradients in the near-wall region.
Estimates for high-Reynolds-number wall-resolved LES indicate that the inner layer can account for most of the grid points \citep{PiomelliBalaras2002}.
Wall-modelled LES (WMLES) alleviates this bottleneck by coupling a near-wall model with an outer-region LES, thereby reducing grid requirements by orders of magnitude \citep{BosePark2018}.
The accuracy of WMLES depends not only on the wall-stress closure itself but also on its coupling to the outer LES, including the choice of matching height and grid resolution \citep{KawaiLarsson2012,HuYangPark2024}.
Recent developments have further highlighted the importance of maintaining the near-wall total-shear-stress balance in mitigating log-layer mismatch \citep{LiuXuHuang2025}, as well as the potential of physics-informed, data-assisted corrections for non-equilibrium near-wall effects \citep{ZhangZhouYangHe2025}.
WMLES of high-Reynolds-number neutral atmospheric boundary layers over homogeneous and heterogeneous rough surfaces have also shown that coupled wall-stress and subgrid-scale treatments can capture the wall shear stress, logarithmic mean-velocity profiles and turbulence spectra across different surface conditions \citep{LiuPullinChengLuo2026}.

Earlier temperature wall-function studies addressed mixed convection by blending forced- and natural-convection profiles in vertical channels \citep{BalajiHoellingHerwig2008} and by evaluating a modified forced-convection wall function in steady Reynolds-averaged simulations around a heated cube \citep{DefraeyeBlockenCarmeliet2012}.
Thermal wall treatments based on the sublayer-resistance formulation of \citet{Jayatilleke1969} have also been used in LES of thermally stratified urban flows \citep{AliabadiKrayenhoffNazarianChewArmstrongAfshariNorford2017, BoppanaXieCastro2014}.
More generally, WMLES formulations for turbulent heat transfer have incorporated models for both wall shear stress and wall heat flux on strongly under-resolved near-wall meshes \citep{KuwataSuga2021}.
For convective atmospheric boundary layers, \citet{SaleskyAnderson2018} employed a wall-stress model based on Monin--Obukhov similarity theory (MOST), which introduces stability corrections to the logarithmic wall law.
\citet{WangYangOvchinnikov2024} assessed LES wall modelling for Rayleigh--B\'enard convection using both \emph{a priori} and \emph{a posteriori} tests of surface shear stress and heat flux, and demonstrated limitations of conventional Monin--Obukhov-based treatments under natural-convection conditions.
These approaches provide useful treatments of buoyancy effects on near-wall momentum and heat transfer, but they were not developed specifically for the finite-height, two-wall PRB configuration.
\citet{ScagliariniEinarssonGylfasonToschi2015} derived a buoyancy-modified law of the wall for unstably stratified turbulent channel flow.
Their formulation provides an important reference for PRB flow, but it is derived for the logarithmic region using asymptotic and local-equilibrium assumptions.
As discussed below, these assumptions can become restrictive under strongly buoyancy-dominated conditions and at finite friction Reynolds numbers, where the mean velocity profile departs substantially from classical logarithmic scaling.
We therefore seek a wall model for incompressible PRB flow that couples momentum and heat transfer and represents buoyancy-induced changes in the mean velocity profile.
Accurate predictions of the wall stress and heat flux are particularly important because errors in the near-wall transport can alter the kinetic-energy distribution and the organisation of large-scale motions in the outer layer.

In this work, we develop a buoyancy-modified logarithmic-quadratic wall model for the canonical PRB configuration.
Its functional form and empirical parameters are assessed using the available PRB DNS database before application in WMLES up to $Re_\tau\approx6000$ and $Ra=10^{10}$.
These simulations examine the spatial organisation and scale-dependent signatures of very-large-scale motions (VLSMs) and buoyancy-driven thermal structures.
The remainder of this paper is organised as follows.
In \S \ref{sec:wall-model-formulation}, we introduce the mixing-length-inspired mean-gradient model and assess its functional form \emph{a priori} against the available DNS data and two reference models.
In \S \ref{sec:numerical-methodology-and-validation}, we describe the computational setup, numerical methods, and the \emph{a posteriori} validation of the WMLES framework.
In \S \ref{sec:turbulent-structures}, we analyse the flow structures and their spectral signatures across the investigated regimes.
In \S \ref{sec:conclusion}, we summarise the main findings of the present work.

\section{Formulation and \emph{a priori} assessment of the buoyancy-modified wall model}
\label{sec:wall-model-formulation}

\subsection{Formulation of the logarithmic-quadratic wall model}
\label{sec:wall-model-derivation}

\begin{figure}
	\centerline{\includegraphics[width=1\textwidth]{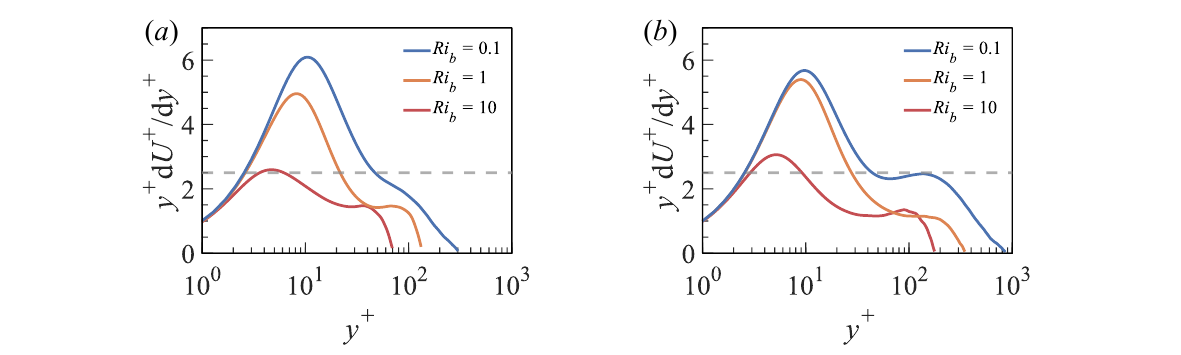}}
	\caption{Logarithmic-law diagnostic function for the PRB system computed from the DNS data of \citet{PirozzoliBernardiniVerziccoOrlandi2017} for (\emph{a}) $Ra=10^7$ and (\emph{b}) $Ra=10^8$.
		The horizontal dashed line marks the classical logarithmic-law value $1/\kappa=2.5$, corresponding to $\kappa=0.4$.}
	\label{fig:log-law-diagnostic}
\end{figure}

In purely shear-driven channel flow, the near-wall velocity follows a linear law in the viscous sublayer and a logarithmic law in the logarithmic region.
Under unstable stratification, with a hot lower wall and a cold upper wall, temperature gradients generate density differences that drive upward and downward motions.
These motions couple the velocity and temperature fields, causing the mean velocity to deviate from the classical wall law.
To quantify this deviation and assess the applicability of the classical scaling, we examine DNS data for the PRB system from \citet{PirozzoliBernardiniVerziccoOrlandi2017} using the logarithmic-law diagnostic function $y^+\mathrm{d}U^+/\mathrm{d}y^+$.
The DNS data considered here correspond to $Pr=\nu/\alpha=1$, where $\nu$ and $\alpha$ are the kinematic viscosity and thermal diffusivity.
The strength of buoyancy relative to shear is measured by the bulk Richardson number $Ri_b=Ra/(Re_b^2Pr)$, where $Ra$ is the Rayleigh number and $Re_b=2hu_b/\nu$ is the Reynolds number based on the full channel height $2h$ and bulk velocity $u_b$.
For a classical logarithmic velocity profile, this diagnostic function is constant at $1/\kappa$, giving a plateau at $2.5$ for $\kappa=0.4$.
Since finite-Reynolds-number effects can also influence the development of this plateau \citep{PirozzoliBernardiniOrlandi2016,MonkewitzNagib2023}, we consider how the diagnostic profiles vary with $Ri_b$ when assessing departures from the classical scaling.
These $Ri_b$-dependent variations, shown in figure \ref{fig:log-law-diagnostic}, therefore motivate a buoyancy-dependent correction to the classical wall law over the near-wall range considered here.
Note that mean temperature--velocity relations have recently been developed for thermal wall modelling in compressible laminar and turbulent wall flows \citep{ChenGanFu2025}.
For passive-scalar transport in incompressible wall-bounded turbulence, \citet{SunFu2026} developed a mean-temperature model based on the relationship between momentum and thermal eddy diffusivities over a wide range of Prandtl numbers.
These formulations concern thermal-transport settings distinct from the present unstably stratified PRB flow, in which temperature is an active scalar that modifies the momentum field through buoyancy.

\begin{figure}
	\centerline{\includegraphics[width=1\textwidth]{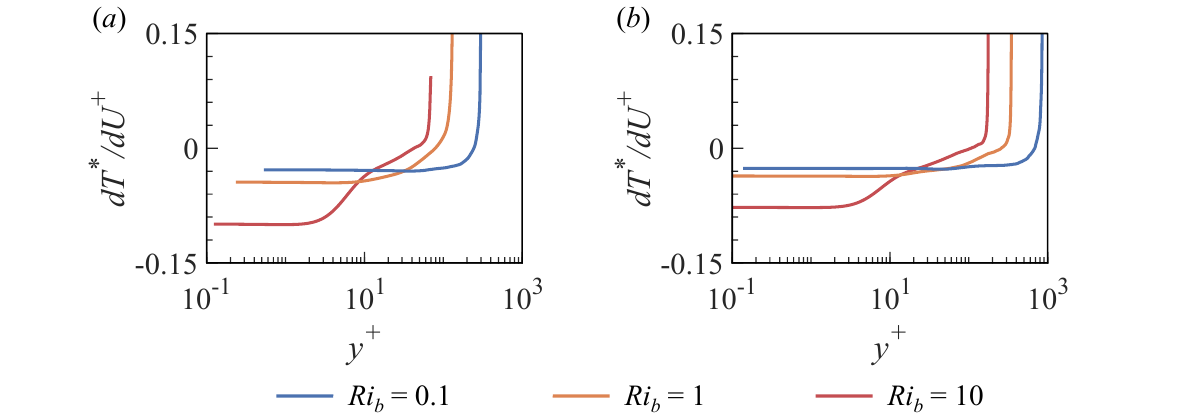}}
	\caption{Wall-normal variation of the local dimensionless temperature--velocity slope defined in equation (\ref{eq:temperature-velocity-slope}), at (\emph{a}) $Ra=10^7$ and (\emph{b}) $Ra=10^8$.
		The derivatives are computed from the plane- and time-averaged DNS temperature and velocity profiles of \citet{PirozzoliBernardiniVerziccoOrlandi2017}.}
	\label{fig:temperature-velocity-slope}
\end{figure}

To express the thermal correction in terms of the mean velocity, we seek a near-wall relation between the mean temperature and mean streamwise velocity.
We first consider the heated lower wall at $y=0$. 
With $T^*=(T-T_c)/\Delta T$ and $\Delta T=T_h-T_c$, we have $T^*=1$ at the wall.
We define the dimensionless local temperature--velocity slope as

\begin{equation}
	\frac{\mathrm{d}T^*}{\mathrm{d}U^+}
	=
	\frac{\mathrm{d}T^*/\mathrm{d}y^+}
	{\mathrm{d}U^+/\mathrm{d}y^+},
	\label{eq:temperature-velocity-slope}
\end{equation}
where $U^+=U/u_\tau$, $y^+=yu_\tau/\nu$ and $u_\tau=(\tau_w/\rho)^{1/2}$, with $\rho$ the fluid density and $\nu$ the kinematic viscosity.
Here, $U(y)$ and $T(y)$ denote the DNS mean profiles of \citet{PirozzoliBernardiniVerziccoOrlandi2017}, obtained by averaging over the homogeneous streamwise--spanwise planes and over time.
The derivatives in equation (\ref{eq:temperature-velocity-slope}) are evaluated from these mean profiles.
As shown in figure \ref{fig:temperature-velocity-slope}, the local slope varies weakly over a finite near-wall range in the cases examined.
This observation motivates an approximately linear relation between the mean temperature and mean streamwise velocity over this range,

\begin{equation}
	T\simeq a+bU.
	\label{eq:linear-temperature-velocity-relation}
\end{equation}
The coefficients $a$ and $b$ are determined by matching the wall value and the ratio of the wall gradients.
The no-slip and thermal boundary conditions, $U(0)=0$ and $T(0)=T_h$, give $a=T_h$.
Matching the slope at the wall gives

\begin{equation}
	b=\left.\frac{\mathrm{d}T/\mathrm{d}y}{\mathrm{d}U/\mathrm{d}y}\right|_{y=0}
	=-\frac{\mu q_w}{k\tau_w},
	\label{eq:wall-temperature-velocity-slope}
\end{equation}
where $q_w$ and $\tau_w$ denote the mean wall heat-flux and shear-stress magnitudes, respectively; at the heated lower wall,
\[
    q_w=-k\left.\frac{\mathrm{d}T}{\mathrm{d}y}\right|_{y=0}>0,
    \qquad
    \tau_w=\mu\left.\frac{\mathrm{d}U}{\mathrm{d}y}\right|_{y=0}>0.
\]
Here, $\mu$ is the dynamic viscosity and $k$ is the thermal conductivity.
Equation (\ref{eq:linear-temperature-velocity-relation}) therefore becomes

\begin{equation}
	T\simeq T_h-\frac{\mu q_w}{k\tau_w}U.
	\label{eq:dimensional-temperature-velocity-approximation}
\end{equation}
In dimensionless form,

\begin{equation}
	T^*\simeq1+\beta U^+,
	\qquad
	\beta=-\frac{Nu_w}{2Re_\tau},
	\label{eq:dimensionless-temperature-velocity-approximation}
\end{equation}
where
\[
    Nu_w
    =
    -\left.
    \frac{\mathrm{d}T^*}{\mathrm{d}y^*}
    \right|_{y^*=0},
    \qquad
    Re_\tau=\frac{u_\tau h}{\nu},
\]
with $y^*=y/H=y/(2h)$, where $H=2h$ is the full channel height.
The quantity $Nu_w$ is computed from the mean wall temperature gradient.

\begin{figure}
	\centerline{\includegraphics[width=1\textwidth]{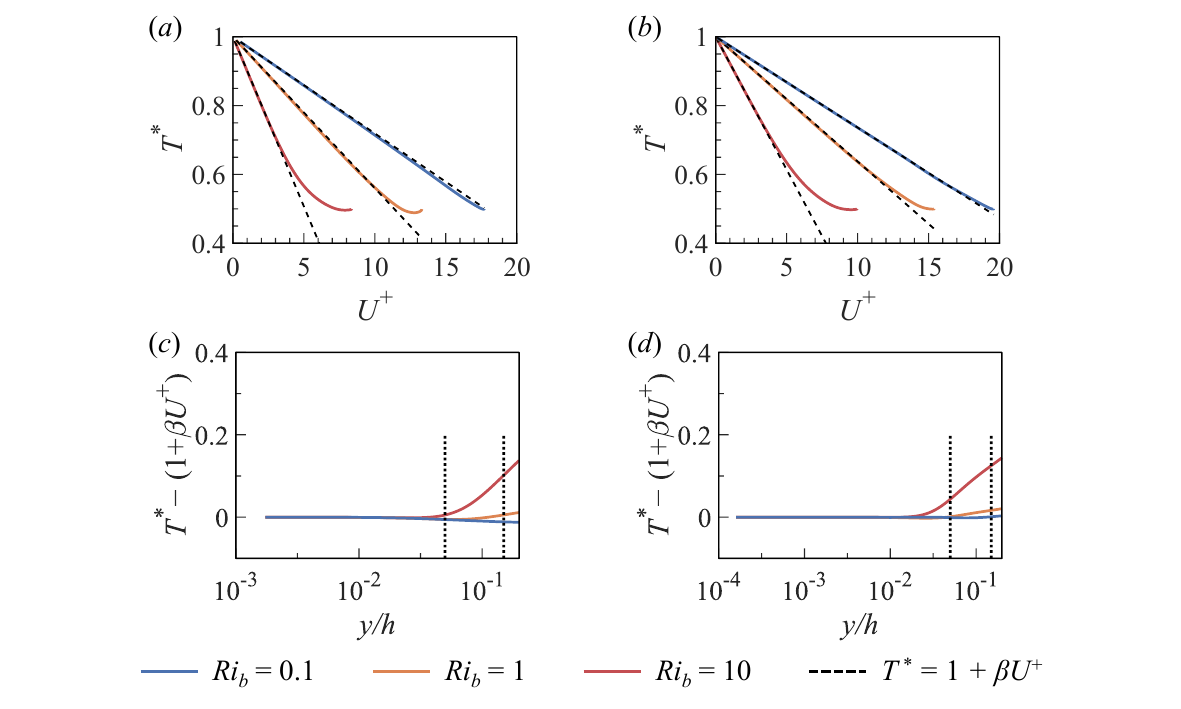}}
	\caption{Assessment of the near-wall linear approximation between the mean temperature and mean streamwise velocity against the DNS data of \citet{PirozzoliBernardiniVerziccoOrlandi2017}.
		(\emph{a},\emph{b}) Comparison of the DNS $T^*$--$U^+$ profiles (solid coloured lines) with the approximation $1+\beta U^+$ in equation (\ref{eq:dimensionless-temperature-velocity-approximation}) (black dashed lines),
		(\emph{c},\emph{d}) the corresponding signed residuals $T^*-(1+\beta U^+)$ versus $y/h$, 
		at (\emph{a},\emph{c}) $Ra=10^7$ and (\emph{b},\emph{d}) $Ra=10^8$.
        The vertical lines in (\emph{c},\emph{d}) mark the wall-model sampling heights $y_p/h$ in the \emph{a priori} comparisons.}
	\label{fig:temperature-velocity-approximation}
\end{figure}

The approximate relation in equation (\ref{eq:dimensionless-temperature-velocity-approximation}) provides the temperature--velocity closure used below to construct the buoyancy-modified momentum wall law.
The upper-wall relation follows from the reflection transformations $d=2h-y$ and $T_u^*=(T_h-T)/\Delta T=1-T^*$.
Using $d$ as the distance measured from the upper wall into the fluid, together with the magnitudes of the wall heat flux and wall shear stress, gives $T_u^*\simeq1+\beta U^+$, with $\beta=-Nu_w/(2Re_\tau)$ and $y^+=u_\tau d/\nu$ evaluated using the upper-wall quantities.
Applying these substitutions to the mean-gradient model below yields the same final velocity wall law at both walls.
Figure \ref{fig:temperature-velocity-approximation}(\emph{a},\emph{b}) compares this near-wall linear temperature--velocity approximation with the DNS mean profiles, using the corresponding DNS wall quantities to determine $\beta$.
To quantify the departure from equation (\ref{eq:dimensionless-temperature-velocity-approximation}), figure \ref{fig:temperature-velocity-approximation}(\emph{c},\emph{d}) shows the residual $r_T(y)=T^*(y)-[1+\beta U^+(y)]$ as a function of $y/h$.
The departures become more pronounced farther from the wall, particularly for $Ri_b=10$.

We next introduce a thermally modified mean-gradient model based on mixing-length scaling.
The temperature--velocity approximation will be used to close this model in terms of the mean velocity.
Continuing with the lower-wall representation, the characteristic velocity and length scales in the neutrally stratified constant-stress layer are $u_m=u_\tau$ and $\ell_m=\kappa y$, respectively.
The corresponding mean velocity gradient is

\begin{equation}
	\frac{\mathrm{d}U}{\mathrm{d}y}
	=
	\frac{u_m}{\ell_m}
	=
	\frac{u_\tau}{\kappa y},
	\label{eq:shear-mean-gradient}
\end{equation}
or, in wall units, $\mathrm{d}U^+/\mathrm{d}y^+=1/(\kappa y^+)$, which is the classical logarithmic velocity gradient.
Under unstable thermal stratification, temperature is an active scalar that modifies turbulent mixing.
Motivated by the mixing-length scaling $\mathrm{d}U/\mathrm{d}y\sim u_m/\ell_m$, we retain the wall-distance-based length scale $\ell_m=\kappa y$ and introduce a phenomenological thermal modulation of the characteristic velocity scale,

\begin{equation}
	u_m=u_\tau\Phi_T\left(T^*\right),
	\qquad
	\Phi_T(1)=1.
	\label{eq:thermal-mixing-velocity}
\end{equation}
This multiplicative form retains $u_\tau$ as the reference velocity determined by the wall stress, while the dimensionless function $\Phi_T$ modulates the characteristic velocity scale according to the local thermal state.
The condition $\Phi_T(1)=1$ normalises the thermal modulation to unity at the hot-wall reference state.

The functional form of $\Phi_T$ is motivated by the classical free-fall velocity scaling used in thermally driven convection.
Under the Boussinesq approximation, the inertial--buoyancy balance $u_f^2/H\sim\beta_T g\Delta T$ gives the global free-fall velocity scale $u_f=\left(\beta_T g\Delta T H\right)^{1/2}$, where $\beta_T$ is the thermal expansion coefficient.
As a modelling assumption, we introduce a temperature-conditioned velocity scale of the form

\begin{equation}
	u_{f,T}(y)=\sqrt{\beta_Tg\left[T(y)-T_c\right]H}.
	\label{eq:local-free-fall-velocity}
\end{equation}
Normalising by $u_f$ gives

\begin{equation}
	\frac{u_{f,T}(y)}{u_f}
	=
	\left[\frac{T(y)-T_c}{\Delta T}\right]^{1/2}
	=
	\left(T^*\right)^{1/2}.
	\label{eq:normalised-free-fall-velocity}
\end{equation}
We adopt the phenomenological thermal modulation $\Phi_T(T^*)=(T^*)^{1/2}$, so that $u_m=u_\tau(T^*)^{1/2}$.
Substituting this thermal modulation into the gradient representation $\mathrm{d}U/\mathrm{d}y=u_m/\ell_m$ with $\ell_m=\kappa y$ gives

\begin{equation}
	\frac{\mathrm{d}U^+}{\mathrm{d}y^+}
	=
	\frac{1}{\kappa y^+}
	\left(T^*\right)^{1/2}.
	\label{eq:thermal-mean-gradient}
\end{equation}
Using the near-wall linear approximation (\ref{eq:dimensionless-temperature-velocity-approximation}), $T^*\simeq1+\beta U^+$, to close equation (\ref{eq:thermal-mean-gradient}) gives the model equation

\begin{equation}
	\frac{\mathrm{d}U^+}{\mathrm{d}y^+}
	=
	\frac{1}{\kappa y^+}
	\left(1+\beta U^+\right)^{1/2}.
	\label{eq:closed-mean-gradient}
\end{equation}

\begin{figure}
	\centerline{\includegraphics[width=1\textwidth]{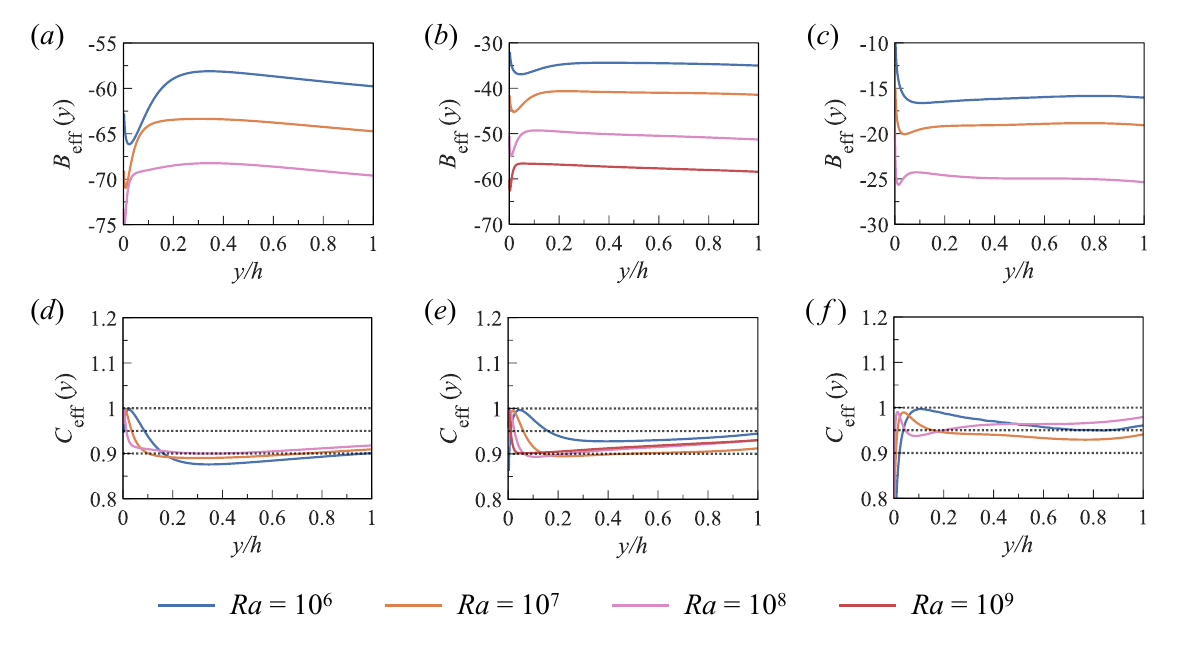}}
	\caption{Pointwise effective parameters inferred from the DNS mean velocity profiles of \citet{PirozzoliBernardiniVerziccoOrlandi2017}.
		(\emph{a}--\emph{c}) $B_{\mathrm{eff}}(y)$ and (\emph{d}--\emph{f}) $C_{\mathrm{eff}}(y)=\beta B_{\mathrm{eff}}(y)/2$ versus wall-normal distance, with (\emph{a},\emph{d}) $Ri_b=0.1$, (\emph{b},\emph{e}) $Ri_b=1$, and (\emph{c},\emph{f}) $Ri_b=10$.
		These pointwise effective values are obtained by rearranging equations (\ref{eq:integrated-velocity-wall-law}) and (\ref{eq:logarithmic-quadratic-wall-law}) using the DNS mean velocity profile $U^+(y^+)$.
		The horizontal dotted lines in (\emph{d}--\emph{f}) indicate $C=0.90$, $0.95$, and $1.00$; the first two correspond to the calibrated values used below, while $C=1.00$ marks the limiting value as $\beta\rightarrow0$.}
	\label{fig:local-integration-parameters}
\end{figure}

For finite $U^+$, the limit $\beta\rightarrow0$ gives $T^*\rightarrow1$ and $\Phi_T\rightarrow1$, so equation (\ref{eq:closed-mean-gradient}) reduces to the classical logarithmic velocity gradient, $\mathrm{d}U^+/\mathrm{d}y^+=1/(\kappa y^+)$.
For $\beta\neq0$, integration of equation (\ref{eq:closed-mean-gradient}) gives

\begin{equation}
	U^+=\frac{\left[\frac{\beta}{2}\left(\frac{1}{\kappa}\ln y^++B\right)\right]^2-1}{\beta},
	\label{eq:integrated-velocity-wall-law}
\end{equation}
where $\kappa=0.4$ is the von K\'arm\'an constant and $B$ is the integration constant for a given case.
The pointwise effective values $B_{\mathrm{eff}}(y)$, inferred from the DNS profiles by inverting equation (\ref{eq:integrated-velocity-wall-law}) at each wall-normal position, exhibit substantial wall-normal variation and also depend strongly on $Ri_b$ and $Ra$, as shown in figure \ref{fig:local-integration-parameters}(\emph{a}--\emph{c}).
This strong dependence complicates the calibration of $B$.
We therefore introduce the rescaled integration parameter $C=\beta B/2$ and rewrite equation (\ref{eq:integrated-velocity-wall-law}) as

\begin{equation}
	U^+=\frac{\left(\frac{\beta}{2\kappa}\ln y^++C\right)^2-1}{\beta}.
	\label{eq:logarithmic-quadratic-wall-law}
\end{equation}
By contrast, the corresponding pointwise effective parameter $C_{\mathrm{eff}}(y)=\beta B_{\mathrm{eff}}(y)/2$ remains close to unity and exhibits substantially smaller absolute wall-normal and $Ra$-dependent variations within the available DNS database (figure \ref{fig:local-integration-parameters}(\emph{d}--\emph{f})).
Note that the profiles $B_{\mathrm{eff}}(y)$ and $C_{\mathrm{eff}}(y)$ describe the parameter values required to match the DNS at each position, whereas the wall law uses one linked pair of casewise constants $B$ and $C$.
In Appendix~\ref{app:integration-parameter-variations}, we quantify the relative variations of the pointwise effective parameters $B_{\mathrm{eff}}(y)$ and $C_{\mathrm{eff}}(y)$ across Rayleigh numbers at fixed wall-normal position.
We also examine how variations in $\beta$ and the wall-normal arithmetic average of $B_{\mathrm{eff}}$ partially offset each other in the corresponding average of $C_{\mathrm{eff}}$.
This reduced variation motivates the use of $C$ in the final formulation.
Because equation (\ref{eq:logarithmic-quadratic-wall-law}) is quadratic in $\ln y^+$, we refer to it as the logarithmic-quadratic wall law.

We next examine the $\beta\rightarrow0$ asymptotic limit of the logarithmic-quadratic wall law.
The reference profile is the classical logarithmic law $U^+=\kappa^{-1}\ln y^++B_{\mathrm{log}}$, where $B_{\mathrm{log}}$ is the additive constant, distinct from the integration constant $B$ in equation (\ref{eq:integrated-velocity-wall-law}).
Expanding equation (\ref{eq:logarithmic-quadratic-wall-law}) separates the constant, logarithmic and quadratic contributions:

\begin{equation}
	U^+=\frac{C^2-1}{\beta}+\frac{C}{\kappa}\ln y^++\frac{\beta}{4\kappa^2}\left(\ln y^+\right)^2.
	\label{eq:expanded-logarithmic-quadratic-wall-law}
\end{equation}
The term quadratic in $\ln y^+$ vanishes as $\beta\rightarrow0$, and recovering the classical logarithmic slope requires $C\rightarrow1$.
The intercept $(C^2-1)/\beta$ also depends on the rate at which $C$ approaches unity.
Matching this intercept to $B_{\mathrm{log}}$ therefore requires
\[
	C=1+\frac{\beta B_{\mathrm{log}}}{2}+o(\beta).
\]
Under this condition, equation (\ref{eq:expanded-logarithmic-quadratic-wall-law}) recovers the classical logarithmic law.
For $\beta=-{Nu}_w/(2{Re}_\tau)<0$ and $B_{\mathrm{log}}>0$, this scaling places $C$ slightly below unity for sufficiently small $|\beta|$.
The finite-stratification values of $C$ used below are instead calibrated over the investigated PRB cases.

\subsection{A priori assessment against DNS}

\begin{figure}
	\centerline{\includegraphics[width=1\textwidth]{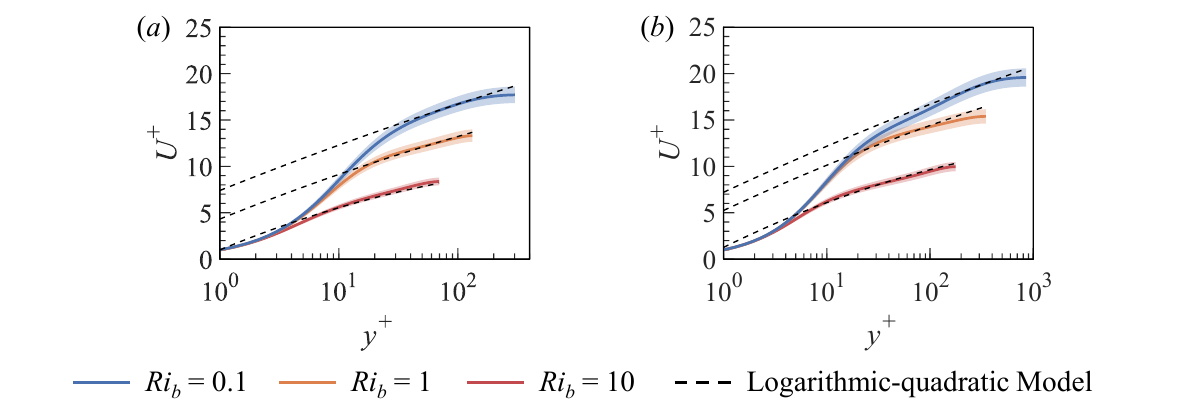}}%
	\caption{\emph{A priori} assessment of the logarithmic-quadratic law against the DNS data of \citet{PirozzoliBernardiniVerziccoOrlandi2017} for (\emph{a}) $Ra=10^7$ and (\emph{b}) $Ra=10^8$.
		The shaded bands indicate pointwise deviations of $\pm5\%$ from the DNS profiles.
		The model uses $\kappa=0.4$, with $C=0.90$ for $Ri_b=0.1$ and $1$, and $C=0.95$ for $Ri_b=10$.} \label{fig:logarithmic-quadratic-law-assessment}
\end{figure}

We first compare the calibrated logarithmic-quadratic profiles with the DNS mean streamwise velocity profiles, as shown in figure \ref{fig:logarithmic-quadratic-law-assessment}.
Across each fixed-$Ra$ sequence, increasing $Ri_b$ is accompanied by flatter profiles and a lower centreline velocity in friction units, $U_c^+=U(h)/u_\tau$.
At fixed $Ra$ and $Pr=1$, increasing $Ri_b$ also reduces $Re_b$; these comparisons therefore do not isolate the response of the dimensional centreline velocity to buoyancy at fixed bulk Reynolds number.
The quadratic dependence on $\ln y^+$ in equation (\ref{eq:logarithmic-quadratic-wall-law}) allows the slope $\mathrm{d}U^+/\mathrm{d}\ln y^+$ to vary with wall-normal distance, providing a correction to the constant slope of the classical logarithmic law.
The comparisons at $Ra=10^7$ and $10^8$ assess whether this functional form can describe the changes in profile shape across the different $Ri_b$ regimes.
Here, we use $C=0.90$ for the investigated $Ri_b=0.1$ and $1$ cases, and $C=0.95$ for $Ri_b=10$, with both values obtained by offline calibration against the available PRB DNS database.

\begin{figure}
	\centerline{\includegraphics[width=1\textwidth]{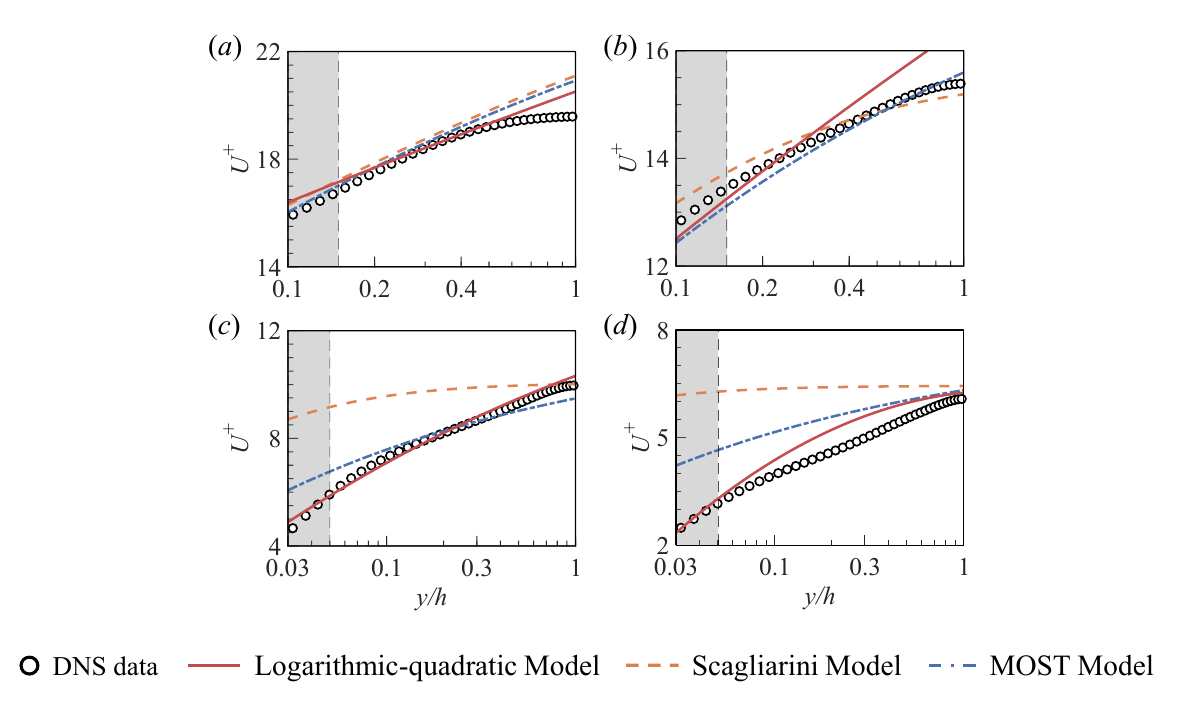}}
	\caption{\emph{A priori} comparison of streamwise-velocity profiles $U^+$ reconstructed using three wall-model functional forms with DNS data of \citet{PirozzoliBernardiniVerziccoOrlandi2017}.
		The MOST curves use the Businger--Dyer implementation in equation (\ref{eq:most-velocity-profile}).
		The DNS-calibrated parameters span $C_1\in[5.8,9.0]$ for Scagliarini's model and $C\in[0.9,1.0]$ for the proposed logarithmic-quadratic model.
		The fitted MOST intercepts are $y_0^+=0.27$, $0.23$, $0.22$ and $0.07$ in (\emph{a})--(\emph{d}), respectively.
        (\emph{a}) $Re_b=10^{4.5}$, corresponding to $Ri_b=0.1$; (\emph{b}) $Re_b=10^{4}$, corresponding to $Ri_b=1$; (\emph{c}) $Re_b=10^{3.5}$, corresponding to $Ri_b=10$; and (\emph{d}) $Re_b=10^{3}$, corresponding to $Ri_b=100$, at $Ra=10^8$.
		The shaded regions indicate the near-wall matching ranges considered in the \emph{a priori} comparisons.}\label{fig:wall-model-velocity-comparison}
\end{figure}

We compare how three DNS-calibrated functional forms reconstruct the mean velocity profile \emph{a priori}: the proposed model, the model of \citet{ScagliariniEinarssonGylfasonToschi2015}, and the Businger--Dyer MOST implementation specified below.
The proposed profile is given by equation (\ref{eq:logarithmic-quadratic-wall-law}), with $\beta=-{Nu_w}/({2Re_\tau})$ evaluated from DNS wall quantities and $C$ calibrated against the same DNS database.
For comparison, the Scagliarini model expresses the velocity profile as

\begin{equation}
	U^+\left(y^+\right)=\frac{1}{\kappa}\ln{\left(\frac{y^+}{1+\kappa_C\,y^+}\right)}+C_1,
	\label{eq:scagliarini-wall-law}
\end{equation}
where $\kappa_C\left(Ra,Re_\tau\right)=A_{\mathrm{Sc}}Ra/(Re_\tau^4Pr^2)$ is the model's buoyancy coefficient, $\kappa=0.42$, and $A_{\mathrm{Sc}}=2.5$ is its fixed empirical constant.
The additive constant $C_1$ is determined separately for each case from the corresponding DNS mean-velocity profile.
For the MOST comparison in the lower-wall representation, we implement the following mean-velocity profile with the integrated stability correction $\mathrm{\Psi}_m$:

\begin{equation}
	U\left(y\right)=\frac{u_\tau}{\kappa}
	\left[
		\ln{\left(\frac{y}{y_0}\right)}
		-\mathrm{\Psi}_m\left(\frac{y}{L_O}\right)
		\right],
	\label{eq:most-velocity-profile}
\end{equation}
where $y_0>0$ is an effective intercept parameter, expressed as an equivalent momentum roughness length, and $\zeta=y/L_O$ is the dimensionless stability parameter.
In the present smooth-wall PRB flow, $y_0$ controls the logarithmic intercept.
We use the positive Obukhov scale $L_O=u_\tau^3/(\beta_TgQ)>0$, where $Q$ is the global kinematic vertical temperature flux.
The corresponding dimensional heat flux $q$ is defined as positive in the upward ($+y$) direction, with $Q=q/(\rho_0c_p)$ and $\alpha=k/(\rho_0c_p)$, where $\rho_0$ is the reference density, $c_p$ is the specific heat capacity at constant pressure, and $k$ is the thermal conductivity.
The global kinematic vertical temperature flux $Q$ used in the Obukhov scale is obtained from the reported global Nusselt number as
\begin{equation}
    Q=\frac{\alpha\Delta T}{2h}Nu,
    \qquad
    q=\rho_0c_pQ=\frac{k\Delta T}{2h}Nu.
    \label{eq:global-temperature-flux}
\end{equation}
For the \emph{a priori} MOST profiles, the corresponding DNS value of this global $Nu$ is used.
For the present unstable flows, the signed global heat transport gives $Q>0$, so that $\zeta>0$ denotes unstable stratification in this implementation.
The stability correction is given by

\[
	\mathrm{\Psi}_m\left(\zeta\right)
	=
	2\ln{\left(\frac{1+x}{2}\right)}
	+\ln{\left(\frac{1+x^2}{2}\right)}
	-2\arctan{\left(x\right)}
	+\frac{\pi}{2},
\]
with $x=\left(1+16\zeta\right)^{1/4}$ and $\kappa=0.35$.
The positive definition of $L_O$, the value $\kappa=0.35$ and the Businger--Dyer coefficient $16$ follow the convention used by \citet{PirozzoliBernardiniVerziccoOrlandi2017} for PRB flow.
The relation adopted here is the integrated Businger--Dyer momentum flux--profile relation.
The underlying gradient relation originates from \citet{BusingerWyngaardIzumiBradley1971} and \citet{Dyer1974}, while the integrated form of the stability correction used here follows \citet{Paulson1970}.
An overview of Monin--Obukhov similarity formulations is provided by \citet{Foken2006}.
For the \emph{a priori} comparison in figure~\ref{fig:wall-model-velocity-comparison}, $y_0$ is fitted separately for each case using the corresponding DNS mean-velocity profile.
Any constant associated with the lower integration limit is absorbed into the fitted effective intercept $y_0$.
The MOST comparisons below refer specifically to the implementation in equation (\ref{eq:most-velocity-profile}).

\begin{figure}
	\centerline{\includegraphics[width=1\textwidth]{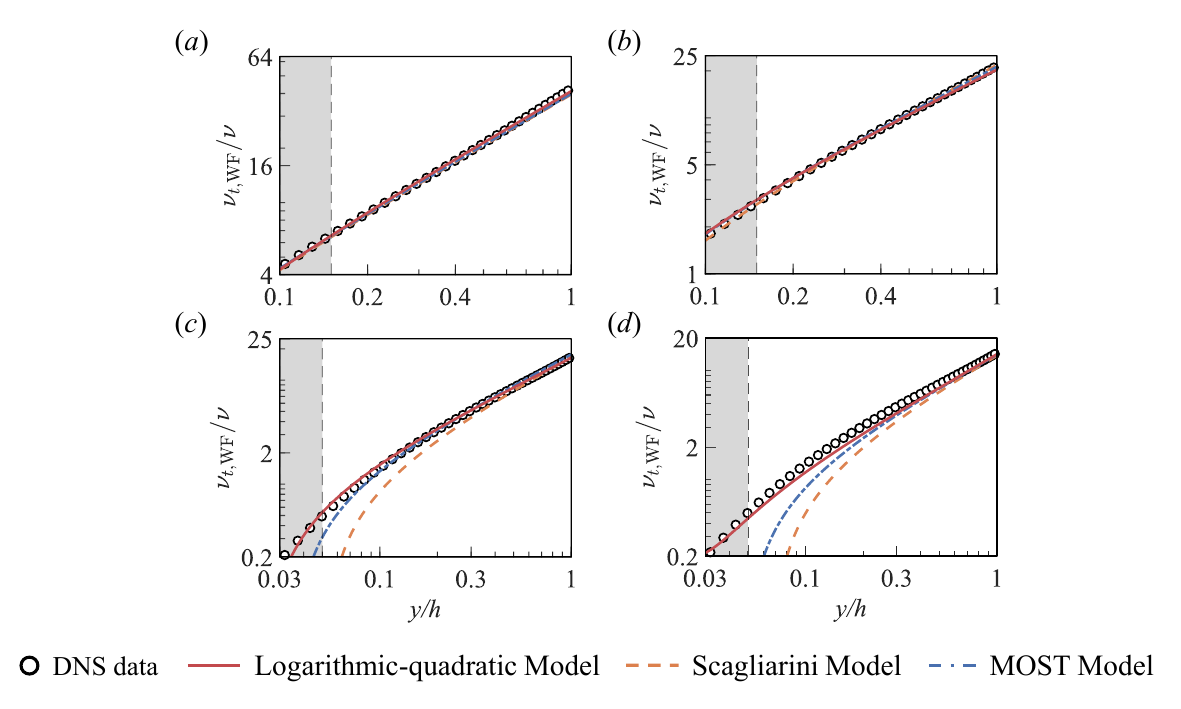}}
	\caption{\emph{A priori} comparison of the wall-function-equivalent eddy viscosity $\nu_{t,\mathrm{WF}}/\nu=y^+/U^+-1$ from the three wall-model formulations and from the DNS mean profiles of \citet{PirozzoliBernardiniVerziccoOrlandi2017}.
		The MOST curves use the Businger--Dyer implementation in equation (\ref{eq:most-velocity-profile}).
		The velocity profiles use the same DNS-calibrated parameters as in figure~\ref{fig:wall-model-velocity-comparison}.
		(\emph{a}) $Re_b=10^{4.5}$, corresponding to $Ri_b=0.1$; (\emph{b}) $Re_b=10^{4}$, corresponding to $Ri_b=1$; (\emph{c}) $Re_b=10^{3.5}$, corresponding to $Ri_b=10$; and (\emph{d}) $Re_b=10^{3}$, corresponding to $Ri_b=100$, at $Ra=10^8$.
		The shaded regions indicate the near-wall matching ranges considered in the \emph{a priori} comparisons.}\label{fig:wall-function-eddy-viscosity}
\end{figure}

As shown in figure \ref{fig:wall-model-velocity-comparison}, all three formulations agree reasonably well with the DNS data in the shear-dominated regime ($Ri_b=0.1$).
Under stronger buoyancy, particularly at $Ri_b=10$ and $100$, the implemented Businger--Dyer MOST profile and the Scagliarini model overpredict the mean velocity over the near-wall matching ranges indicated in figure \ref{fig:wall-model-velocity-comparison}, whereas the proposed functional form provides a closer reconstruction of the DNS profiles.
We also express the same profile comparison in terms of the wall-function-equivalent eddy viscosity,
\begin{equation}
    \frac{\nu_{t,\mathrm{WF}}}{\nu}=\frac{y^+}{U^+}-1,
\end{equation}
obtained by representing the wall shear stress magnitude as $\tau_w=\rho\left(\nu+\nu_{t,\mathrm{WF}}\right)U/y$.
Here, $U/y$ is the wall-to-point velocity gradient magnitude for the lower-wall mean profile, distinct from the local gradient $\mathrm{d}U/\mathrm{d}y$.
Figure \ref{fig:wall-function-eddy-viscosity} shows that the implemented Businger--Dyer MOST profile and the Scagliarini model underestimate this quantity in the matching region under strong buoyancy, whereas the proposed model gives a closer reconstruction.

The differences in profile reconstruction may reflect the models' assumptions and calibration.
First, the model of \citet{ScagliariniEinarssonGylfasonToschi2015} was derived specifically for the logarithmic region.
Its derivation uses the asymptotic $Re_\tau\rightarrow\infty$ form of the total-stress balance.
It further assumes an approximately constant Reynolds stress, local equilibrium between turbulent-energy production and dissipation, and a Prandtl mixing length proportional to $y$.
The buoyancy-production term is closed using a local gradient-diffusion approximation for the turbulent heat flux together with an assumed logarithmic mean-temperature gradient.
These assumptions may be restrictive when the DNS velocity and temperature profiles depart substantially from logarithmic behaviour.
In particular, the case $Ri_b=100$ at $Ra=10^8$ has $Re_\tau\approx96$, so a well-developed asymptotic logarithmic region cannot be assumed.
Second, the Businger--Dyer relation underlying the MOST implementation used here was developed primarily from atmospheric constant-flux surface-layer observations.
It assumes that the dimensionless mean-velocity gradient is a universal function of the single local stability parameter $y/L_O$.
In the present PRB system, finite-height and two-wall confinement may introduce an outer-scale dependence and non-local momentum and heat transport not captured by a relation based on $y/L_O$ alone.
Related limitations of Monin--Obukhov-based wall treatments under natural-convection-dominated conditions have also been identified in \emph{a priori} and \emph{a posteriori} LES wall-model assessments of Rayleigh--B\'enard convection \citep{WangYangOvchinnikov2024}.
The proposed model uses a mixing-length-based gradient representation over the specified matching interval, incorporating the near-wall linear temperature--velocity approximation and the DNS wall heat flux through $\beta$.
Together with the calibration of $C$, these inputs may contribute to the improved profile reconstruction in the matching region.

\section{Numerical methodology and \emph{a posteriori} validation}\label{sec:numerical-methodology-and-validation}

\subsection{Computational setup and governing equations}

We consider incompressible flow between two horizontal walls, with a hot lower wall at temperature $T_h$ located at $y=0$ and a cold upper wall at temperature $T_c$ located at $y=2h$.
The flow is driven by a spatially uniform streamwise body-force acceleration $f_b$ representing the mean pressure-gradient forcing.
A constant bulk flow rate is imposed, and $f_b$ is adjusted dynamically at each time step to maintain the prescribed bulk Reynolds number $Re_b$.
Periodic boundary conditions are imposed in the streamwise and spanwise directions, and no-slip velocity conditions are applied at both walls.
The computational domain is $L\times H\times W=16h\times 2h\times 8h$ in the streamwise ($x$), wall-normal ($y$), and spanwise ($z$) directions, respectively.
The flow is characterised by three independent dimensionless numbers: the bulk Reynolds number, $Re_b=2hu_b/\nu$, where $u_b$ is the bulk velocity; the Rayleigh number, $Ra=(8h^3\beta_T g\Delta T)/(\alpha\nu)$; and the Prandtl number, $Pr=\nu/\alpha$.
The strength of buoyancy relative to shear is quantified by the bulk Richardson number $Ri_b=Ra/(Re_b^2Pr)$.
The main parameter study spans $10^8 \le Ra \le 10^{10}$ and $0.1 \le Ri_b \le 10$, with $Pr=1$ fixed.
All simulations are performed using OpenFOAM (version 8) with the transient \emph{buoyantPimpleFoam} solver under the Boussinesq approximation and a subgrid-scale turbulence model.
The filtered governing equations are written in component form as

\begin{subequations}\label{eq:dimensional-governing-equations}
    \begin{align}
        \frac{\partial\overline{u_i}}{\partial x_i}&=0,\label{eq:dimensional-continuity}\\
        \frac{\partial\overline{u_i}}{\partial t}
        +\frac{\partial\left(\overline{u_i}\,\overline{u_j}\right)}{\partial x_j}
        &=-\frac{1}{\rho_0}\frac{\partial\overline{p}}{\partial x_i}
        +\nu\frac{\partial^2\overline{u_i}}{\partial x_j\partial x_j}
        -\frac{\partial\tau_{ij}^{\mathrm{sgs}}}{\partial x_j}
        +g\beta_T\left(\overline{T}-T_0\right)\delta_{i2}
        +f_b\delta_{i1},\label{eq:dimensional-momentum}\\
        \frac{\partial\overline{T}}{\partial t}
        +\frac{\partial\left(\overline{T}\,\overline{u_i}\right)}{\partial x_i}
        &=\alpha\frac{\partial^2\overline{T}}{\partial x_i\partial x_i}
        -\frac{\partial J_i^{\mathrm{sgs}}}{\partial x_i}.\label{eq:dimensional-temperature}
    \end{align}
\end{subequations}
The SGS kinematic stress and temperature flux are defined by $\tau_{ij}^{\mathrm{sgs}}=\overline{u_i u_j}-\overline{u_i}\,\overline{u_j}$ and $J_i^{\mathrm{sgs}}=\overline{T u_i}-\overline{T}\,\overline{u_i}$.
Here, the overbar denotes spatial filtering, repeated indices are summed, and $\delta_{ij}$ is the Kronecker delta, with $x_1=x$, $x_2=y$ and $x_3=z$; $\rho_0$ is the reference density, $g$ is the magnitude of the gravitational acceleration, and $T_0=T_c$ is the reference temperature in the Boussinesq approximation.

We introduce the nondimensional variables
\begin{equation}
\begin{aligned}
x_i^* &= \frac{x_i}{2h},
&
t^* &= \frac{t}{2h/u_b},
&
u_i^* &= \frac{u_i}{u_b},
\\
p^* &= \frac{p}{\rho_0u_b^2},
&
T^* &= \frac{T-T_c}{\Delta T},
&
f_b^* &= \frac{f_b}{u_b^2/(2h)},
\\
\tau_{ij}^{\mathrm{sgs}*}
&= \frac{\tau_{ij}^{\mathrm{sgs}}}{u_b^2},
&
J_i^{\mathrm{sgs}*}
&= \frac{J_i^{\mathrm{sgs}}}{u_b\Delta T}.
\end{aligned}
\label{eq:nondimensional_variables}
\end{equation}
Equations (\ref{eq:dimensional-governing-equations}) can then be written in nondimensional form as

\begin{subequations}\label{eq:nondimensional-governing-equations}
    \begin{align}
        \frac{\partial\overline{u_i^*}}{\partial x_i^*}&=0,\label{eq:nondimensional-continuity}\\
        \frac{\partial\overline{u_i^*}}{\partial t^*}
        +\frac{\partial\left(\overline{u_i^*}\,\overline{u_j^*}\right)}{\partial x_j^*}
        &=-\frac{\partial\overline{p^*}}{\partial x_i^*}
        +\frac{1}{Re_b}\frac{\partial^2\overline{u_i^*}}{\partial x_j^*\partial x_j^*}
        -\frac{\partial\tau_{ij}^{\mathrm{sgs}*}}{\partial x_j^*}+\frac{Ra}{Re_b^2Pr}\overline{T^*}\delta_{i2}
        +f_b^*\delta_{i1},\label{eq:nondimensional-momentum}\\
        \frac{\partial\overline{T^*}}{\partial t^*}
        +\frac{\partial\left(\overline{T^*}\,\overline{u_i^*}\right)}{\partial x_i^*}
        &=\frac{1}{Re_bPr}\frac{\partial^2\overline{T^*}}{\partial x_i^*\partial x_i^*}
        -\frac{\partial J_i^{\mathrm{sgs}*}}{\partial x_i^*}.\label{eq:nondimensional-temperature}
    \end{align}
\end{subequations}
For the velocity field, we use the wall-adapting local eddy-viscosity (WALE) model \citep{NicoudDucros1999}, while for the temperature field we adopt a constant SGS Prandtl number.
The WALE model accounts for strain-rate and rotation-rate contributions through the velocity-gradient tensor.
Dynamic and mixed-scale alternatives have been studied for buoyancy-driven turbulence \citep{LauYeohTimchenkoReizes2013,Yilmaz2021,DabbaghTriasGorobetsOliva2017,MaulikSan2017}.
Here we use the fixed algebraic WALE closure for computational robustness; the resolved statistics and global heat-transfer performance are assessed below.

\begin{figure}
	\centerline{\includegraphics[width=1\textwidth]{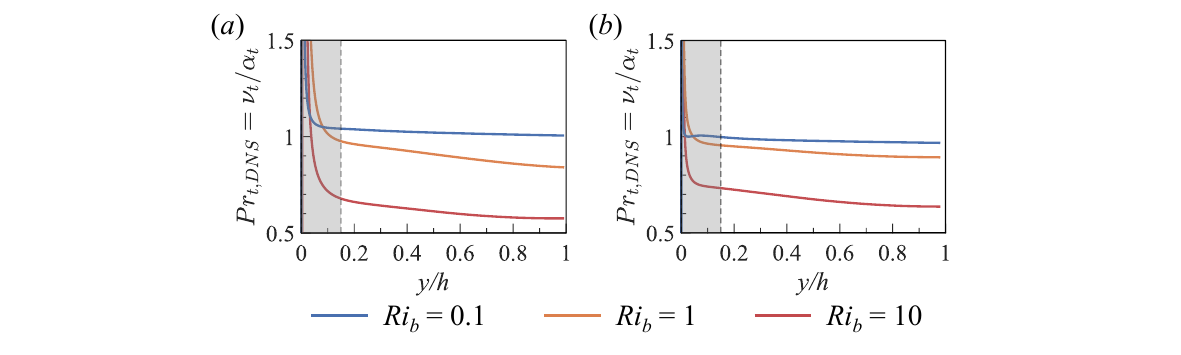}}
	\caption{Wall-normal variation of the DNS-inferred turbulent Prandtl ratio $Pr_{t,\mathrm{DNS}}=\nu_{t,\mathrm{DNS}}/\alpha_{t,\mathrm{DNS}}$, obtained from the DNS data of \citet{PirozzoliBernardiniVerziccoOrlandi2017}, for (\emph{a}) $Ra=10^7$ and (\emph{b}) $Ra=10^8$.
		The shaded region indicates the range of wall-adjacent cell-centre locations used as the wall-model sampling heights across the WMLES cases.}\label{fig:wall-turbulent-prandtl-number}
\end{figure}

The framework uses separate turbulent Prandtl numbers: $Pr_{\mathrm{sgs}}$ for the SGS heat-flux closure and $Pr_{t,\mathrm{wall}}$ for the thermal wall function.
These parameters describe different modelled transport contributions and are prescribed independently.
The bulk SGS eddy viscosity and thermal diffusivity are denoted by $\nu_{\mathrm{sgs}}$ and $\alpha_{\mathrm{sgs}}$, respectively, with $\alpha_{\mathrm{sgs}}=\nu_{\mathrm{sgs}}/Pr_{\mathrm{sgs}}$, and we set $Pr_{\mathrm{sgs}}=0.4$ \citep{KenjeresHanjalic2006}.
For the DNS mean profiles, we define the flux--gradient transport coefficients as
\[
    \nu_{t,\mathrm{DNS}}=-\frac{\langle u'v'\rangle_{x,z,t}}{\mathrm{d}U/\mathrm{d}y},
    \qquad
    \alpha_{t,\mathrm{DNS}}=-\frac{\langle v'T'\rangle_{x,z,t}}{\mathrm{d}T/\mathrm{d}y},
\]
where $\langle\cdot\rangle_{x,z,t}$ denotes plane and time averaging.
We use the DNS-inferred ratio $Pr_{t,\mathrm{DNS}}=\nu_{t,\mathrm{DNS}}/\alpha_{t,\mathrm{DNS}}$, obtained from the data of \citet{PirozzoliBernardiniVerziccoOrlandi2017} and shown in figure \ref{fig:wall-turbulent-prandtl-number}, to select representative values of the prescribed wall turbulent Prandtl number $Pr_{t,\mathrm{wall}}$.
Very close to the wall, both turbulent transport coefficients approach zero and their ratio is ill-conditioned.
We instead use the approximately constant portions of the profiles farther from the wall.
Across the available DNS range shown in figure \ref{fig:wall-turbulent-prandtl-number}, the representative plateau values of $Pr_{t,\mathrm{DNS}}$ vary only weakly with $Ra$ at fixed $Ri_b$. 
We therefore prescribe $Pr_{t,\mathrm{wall}} = 0.9$ for $Ri_b = 0.1$ and $1$ and $Pr_{t,\mathrm{wall}} = 0.7$ for $Ri_b = 10$. 
These values are retained at higher $Ra$ as a modelling assumption; any additional $Ra$-dependence outside the available DNS range constitutes an uncertainty of the high-$Ra$ extrapolation.
The shaded region in figure \ref{fig:wall-turbulent-prandtl-number} indicates the range of wall-adjacent cell-centre locations used as the wall-model sampling heights across the WMLES cases.

In all simulations, temporal discretisation uses OpenFOAM's second-order implicit backward scheme.
The time step is adjusted to maintain a maximum Courant--Friedrichs--Lewy (CFL) number of $0.8$.
In each wall-normal column, the control volume adjacent to either wall has a thickness $2y_p$, so that its centroid is located at the prescribed sampling height $y_p$ measured from the wall boundary face.
The two wall-adjacent cells are symmetric about the channel centreline.
The remaining $N_y-2$ cells are uniformly spaced in the wall-normal direction, with interior spacing $\Delta y_{\mathrm{int}}=(2h-4y_p)/(N_y-2)$.
The wall-adjacent cell centres form homogeneous ($x$--$z$) planes at the wall-model sampling height $y_p$.
All statistics are collected over $500t_f$ after an initial $500t_f$ transient, where $t_f=H/u_f=\sqrt{H/(g\beta_T\Delta T)}$ is the free-fall time.
At each time step, the wall-model inputs are obtained by averaging the resolved cell-centred fields over the homogeneous sampling plane, separately for the lower and upper walls.
For the lower wall, the friction velocity is evaluated using the time-averaged inputs $\overline U_p=\langle U_p(t)\rangle_t=\langle u_x(x,y_p,z,t)\rangle_{x,z,t}$ and $\overline{\nu}_{t,w}=\langle\nu_{t,w}(t)\rangle_t$ as
\begin{equation}
    u_\tau=
    \left[
        \left(\nu+\overline{\nu}_{t,w}\right)
        \frac{\overline U_p}{y_p}
    \right]^{1/2}.
    \label{eq:statistical-heated-wall-friction-velocity}
\end{equation}
This value is used for the reported friction-based quantities and normalisations.
Table \ref{tab:simulation-parameters} lists the simulation parameters and grid settings, and table \ref{tab:grid-resolution} gives the corresponding spacings in wall units, $\Delta x^+$, $\Delta y_{\mathrm{int}}^+$ and $\Delta z^+$.
The wall model represents unresolved near-wall transport, the outer mesh is intended to resolve energy-containing motions, and the DNS comparisons below assess the resulting statistics.
The matching height and grid resolution can influence both the mean-flow accuracy and the convergence behaviour of WMLES \citep{KawaiLarsson2012,HuYangPark2024}.
We therefore assess the adopted grid and matching-height choices \emph{a posteriori} against the available PRB DNS and compare the corresponding spacings with representative WMLES resolutions reported in the literature.
For the most demanding case, $Ra=10^{10}$ and $Ri_b=0.1$ ($Re_\tau\approx6000$), the spacings in the channel interior are $\Delta x^+\approx128.0$, $\Delta y_{\mathrm{int}}^+\approx64.0$ and $\Delta z^+\approx64.0$.
For reference, \citet{BaeLozanoDuranBoseMoin2019} used $\Delta_i^+=210$ for plane-channel WMLES at $Re_\tau\approx4200$; \citet{ChungMcKeon2010} used $(\Delta x^+,\Delta y^+,\Delta z^+)\approx(340,85,340)$ for long-channel WMLES at $Re_\tau=2000$; and \citet{DAlessandroMarchioliPiomelli2025} used $\Delta x^+=156$, $\Delta y^+=56$ and $\Delta z^+=78$ in a channel-flow validation at $Re_\tau\approx5200$.

\begin{table}
	\begin{center}
		\def~{\hphantom{0}}
		\newcommand{\tabstat}[2]{#2}
		\begin{tabular}{lllllllllll}
			Flow case    & $Re_b$ & $Ra$      & $Ri_b$ & $h/L_O$       & $Re_\tau$     & $Nu$          & $C_f$               & $N_x$ & $N_y$ & $N_z$ \\
			Ra7Re3.5     & 3162   & $10^7$    & 1      & \tabstat{2.98641455}{2.99} & \tabstat{129.98}{130} & \tabstat{10.493}{10.5} & $\tabstat{0.01351584032}{1.35\times10^{-2}}$ & 128   & 42    & 128   \\
			Ra7Re3.5$^*$ & 3162   & $10^7$    & 1      & \tabstat{3.01917430}{3.02} & \tabstat{134.98}{135} & \tabstat{11.880}{11.9} & $\tabstat{0.01457568032}{1.46\times10^{-2}}$ & 1024  & 256   & 512   \\

			Ra8Re3.5     & 3162   & $10^8$    & 10     & \tabstat{36.332391}{36.3} & \tabstat{169.44579}{169} & \tabstat{28.281754}{28.3} & $\tabstat{0.022969502}{2.30\times10^{-2}}$ & 256   & 66    & 256   \\
			Ra8Re3.5$^*$ & 3162   & $10^8$    & 10     & \tabstat{30.09458263}{30.1} & \tabstat{179.12}{179} & \tabstat{27.672}{27.7} & $\tabstat{0.02566717952}{2.57\times10^{-2}}$ & 2560  & 512   & 1280  \\

			Ra8Re4       & 10000  & $10^8$    & 1      & \tabstat{4.0522033}{4.05} & \tabstat{337.78215}{338} & \tabstat{24.987414}{25.0} & $\tabstat{0.0091277427}{9.13\times10^{-3}}$ & 256   & 66    & 256   \\
			Ra8Re4$^*$   & 10000  & $10^8$    & 1      & \tabstat{3.67696814}{3.68} & \tabstat{351.01}{351} & \tabstat{25.443}{25.4} & $\tabstat{0.009856641608}{9.86\times10^{-3}}$ & 2560  & 512   & 1280  \\

			Ra8Re4.5     & 31623  & $10^8$    & 0.1    & \tabstat{0.52364729}{0.524} & \tabstat{792.40145}{792} & \tabstat{41.686424}{41.7} & $\tabstat{0.0050232005}{5.02\times10^{-3}}$ & 256   & 66    & 256   \\
			Ra8Re4.5$^*$ & 31623  & $10^8$    & 0.1    & \tabstat{0.44135633}{0.441} & \tabstat{864.24}{864} & \tabstat{45.584}{45.6} & $\tabstat{0.0059752862208}{5.98\times10^{-3}}$ & 2560  & 512   & 1280  \\

			Ra9Re4       & 10000  & $10^9$    & 10     & \tabstat{44.910937}{44.9} & \tabstat{448.85698}{449} & \tabstat{64.982444}{65.0} & $\tabstat{0.016117807}{1.61\times10^{-2}}$ & 512   & 102   & 512   \\

			Ra9Re4.5     & 31623  & $10^9$    & 1      & \tabstat{4.8776761}{4.88} & \tabstat{893.46316}{893} & \tabstat{55.662522}{55.7} & $\tabstat{0.0063862113}{6.39\times10^{-3}}$ & 512   & 102   & 512   \\
			Ra9Re4.5$^*$ & 31623  & $10^9$    & 1      & \tabstat{4.44257887}{4.44} & \tabstat{946.41}{946} & \tabstat{60.255}{60.3} & $\tabstat{0.0071655351048}{7.17\times10^{-3}}$ & 6144  & 768   & 3072  \\

			Ra9Re5       & 100000 & $10^9$    & 0.1    & \tabstat{0.60271844}{0.603} & \tabstat{2180.1562}{2180} & \tabstat{99.930319}{99.9} & $\tabstat{0.0038024648}{3.80\times10^{-3}}$ & 512   & 102   & 512   \\

			Ra10Re4.5    & 31623  & $10^{10}$ & 10     & \tabstat{50.536398}{50.5} & \tabstat{1250.9966}{1250} & \tabstat{158.30428}{158} & $\tabstat{0.012519940}{1.25\times10^{-2}}$ & 750   & 152   & 750   \\

			Ra10Re5      & 100000 & $10^{10}$ & 1      & \tabstat{5.8298414}{5.83} & \tabstat{2462.7946}{2460} & \tabstat{139.33535}{139} & $\tabstat{0.0048522859}{4.85\times10^{-3}}$ & 750   & 152   & 750   \\

			Ra10Re5.5    & 316228 & $10^{10}$ & 0.1    & \tabstat{0.71825440}{0.718} & \tabstat{5999.2531}{6000} & \tabstat{248.13603}{248} & $\tabstat{0.0028792830}{2.88\times10^{-3}}$ & 750   & 152   & 750   \\
		\end{tabular}
		\caption{Numerical details of the simulated flow cases.
			Here, $Re_b=2hu_b/\nu$ is the bulk Reynolds number; $Ra=(8h^3\beta_Tg\Delta T)/(\alpha\nu)$ is the Rayleigh number; $Ri_b=2\beta_Tg\Delta T h/u_b^2$ is the bulk Richardson number; $L_O=u_\tau^3/(\beta_TgQ)>0$ is the positive Obukhov scale, where $Q=\alpha\Delta T Nu/(2h)>0$ is the global kinematic vertical temperature flux obtained from the reported $Nu$; $Re_\tau=hu_\tau/\nu$ is the friction Reynolds number; $Nu$ is the volume- and time-averaged global Nusselt number; and $C_f=2u_\tau^2/u_b^2=8(Re_\tau/Re_b)^2$ is the skin-friction coefficient.
			The derived quantities $h/L_O=Ra\,Nu/(16Pr^2Re_\tau^3)$ and $C_f$ are evaluated using the retained precision of $Nu$ and $Re_\tau$, with $Re_b$ given by the case definition.
			For the WMLES entries, $N_x$, $N_y$, and $N_z$ denote the numbers of control volumes (cells) in the streamwise, wall-normal, and spanwise directions, respectively; the DNS mesh counts are retained as reported in the source study.
			The notation RaXReY denotes configurations with $Ra=10^\mathrm{X}$ and $Re_b=10^\mathrm{Y}$.
			The superscript $^*$ indicates the results of \citet{PirozzoliBernardiniVerziccoOrlandi2017}.}
		\label{tab:simulation-parameters}
	\end{center}
\end{table}

\begin{table}
	\begin{center}
		\def~{\hphantom{0}}
		\begin{tabular}{lllllllll}
			Flow case & $Re_b$ & $Ra$      & $Ri_b$ & $\Delta x^+$  & $\Delta y_{\mathrm{int}}^+$  & $\Delta z^+$  & $y_p^+$   & $y_p/h$    \\
			Ra7Re3.5  & 3162   & $10^7$    & 1      & 16.25         & 4.549         & 8.124         & 19.50     & 0.15       \\
			Ra8Re3.5  & 3162   & $10^8$    & 10     & 10.58         & 4.762         & 5.292         & 8.467     & 0.05       \\
			Ra8Re4    & 10000  & $10^8$    & 1      & 21.08         & 7.377         & 10.54         & 50.58     & 0.15       \\
			Ra8Re4.5  & 31623  & $10^8$    & 0.1    & 49.53         & 17.33         & 24.76         & 118.9     & 0.15       \\
			Ra9Re4    & 10000  & $10^9$    & 10     & 14.12         & 8.132         & 7.059         & 22.59     & 0.05       \\
			Ra9Re4.5  & 31623  & $10^9$    & 1      & 28.63         & 12.83         & 14.31         & 137.4     & 0.15       \\
			Ra9Re5    & 100000 & $10^9$    & 0.1    & 68.13         & 30.52         & 34.06         & 327.0     & 0.15       \\
			Ra10Re4.5 & 31623  & $10^{10}$ & 10     & 27.04         & 15.88         & 13.52         & 38.02     & 0.03       \\
			Ra10Re5   & 100000 & $10^{10}$ & 1      & 54.42         & 27.21         & 27.21         & 255.1     & 0.10       \\
			Ra10Re5.5 & 316228 & $10^{10}$ & 0.1    & 128.0         & 63.99         & 63.99         & 599.9     & 0.10       \\
		\end{tabular}
		\caption{Grid resolutions for the various flow cases.
			Here, $\Delta x^+=u_\tau\Delta x/\nu$, $\Delta y_{\mathrm{int}}^+=u_\tau\Delta y_{\mathrm{int}}/\nu$, $\Delta z^+=u_\tau\Delta z/\nu$ and $y_p^+=u_\tau y_p/\nu$ use the statistical friction velocity from the heated lower wall.
			The interior spacing $\Delta y_{\mathrm{int}}$ is the uniform wall-normal cell spacing, whereas the thickness of each wall-adjacent cell is $2y_p$.
			Here, $y_p$ is the distance from the corresponding wall boundary face to the centroid of the wall-adjacent control volume.
			The wall-adjacent cell centres form the wall-model sampling plane.
			The notation RaXReY denotes configurations with $Ra=10^\mathrm{X}$ and $Re_b=10^\mathrm{Y}$.}
		\label{tab:grid-resolution}
	\end{center}
\end{table}

To couple momentum and heat transfer at the wall, the model uses the instantaneous quantities averaged over the sampling plane in a separate iterative solve for each wall at each time step.
Specifically, at the lower wall we define
\[
    U_p(t)=\langle u_x(x,y_p,z,t)\rangle_{x,z},
    \qquad
    T_p(t)=\langle T(x,y_p,z,t)\rangle_{x,z},
\]
where $\langle\cdot\rangle_{x,z}$ denotes an instantaneous plane average.
At the upper wall, the corresponding sampling-plane quantities are evaluated at $y=2h-y_p$.
For each wall, the iteration determines scalar coefficients $\nu_{t,w}(t)$ and $\alpha_{t,w}(t)$, which are spatially uniform over that wall patch.
The corresponding total effective viscosity and thermal diffusivity include the molecular contributions, namely, $\nu_{\mathrm{eff},w}(t)=\nu+\nu_{t,w}(t)$ and $\alpha_{\mathrm{eff},w}(t)=\alpha+\alpha_{t,w}(t)$.
Given an initial estimate of $\nu_{t,w}(t)$, we obtain the corresponding thermal contribution as $\alpha_{t,w}(t)=\nu_{t,w}(t)/Pr_{t,\mathrm{wall}}$ and evaluate the instantaneous plane-averaged wall-model Nusselt number as
\begin{equation}
    Nu_w(t)=
    \frac{2h}{\Delta T}
    \left(\frac{\alpha_{\mathrm{eff},w}(t)}{\alpha}\right)
    \frac{|T_w-T_p(t)|}{y_p},
    \label{eq:instantaneous-wall-nusselt-number}
\end{equation}
where $T_w=T_h$ on the lower wall and $T_w=T_c$ on the upper wall.
The corresponding instantaneous plane-averaged wall heat-flux magnitude is $q_w(t)=k\Delta T Nu_w(t)/(2h)$.
The resulting $Nu_w(t)$ enters the momentum wall law through $\beta(t)=-Nu_w(t)/(2Re_\tau(t))$, thereby coupling the thermal and momentum wall treatments.
The instantaneous friction velocity is evaluated from the corresponding plane-averaged streamwise wall-shear-stress magnitude as $u_\tau(t)=\sqrt{|\tau_w(t)|/\rho}$, and the corresponding friction Reynolds number is $Re_\tau(t)=hu_\tau(t)/\nu$.

The logarithmic-quadratic wall law in equation (\ref{eq:logarithmic-quadratic-wall-law}) is then solved using the Newton--Raphson method for the dimensionless sampling height $y_p^+(t)=u_\tau(t)y_p/\nu$, whose converged value gives an updated $\nu_{t,w}(t)$.
The thermal update and Newton--Raphson solve are repeated until the relative change in $\nu_{t,w}(t)$ is below $1\%$.
After convergence, the final thermal contribution $\alpha_{t,w}(t)$ is obtained from the converged $\nu_{t,w}(t)$ using the prescribed constant $Pr_{t,\mathrm{wall}}$.
The converged scalar coefficients $\nu_{t,w}(t)$ and $\alpha_{t,w}(t)$ are applied uniformly over the corresponding wall patch.
The velocity satisfies a no-slip condition and the wall temperature is prescribed by a fixed-value condition.
In the OpenFOAM implementation, the turbulent viscosity is supplied through the custom \texttt{nutPRBWallFunction}, and the turbulent thermal diffusivity is treated using the corresponding thermal wall-function boundary condition.
When the momentum and temperature equations are advanced, the local wall-face momentum and heat fluxes are evaluated using these uniform wall coefficients together with the local velocity and temperature gradients.
The same scalar effective viscosity $\nu_{\mathrm{eff},w}(t)$ acts on both tangential velocity components through the viscous term.
The lower and upper walls are treated independently, so their instantaneous wall coefficients and plane-averaged wall fluxes may differ and evolve in time.

\subsection{A posteriori validation against DNS}

\begin{figure}
	\centerline{\includegraphics[width=1\textwidth]{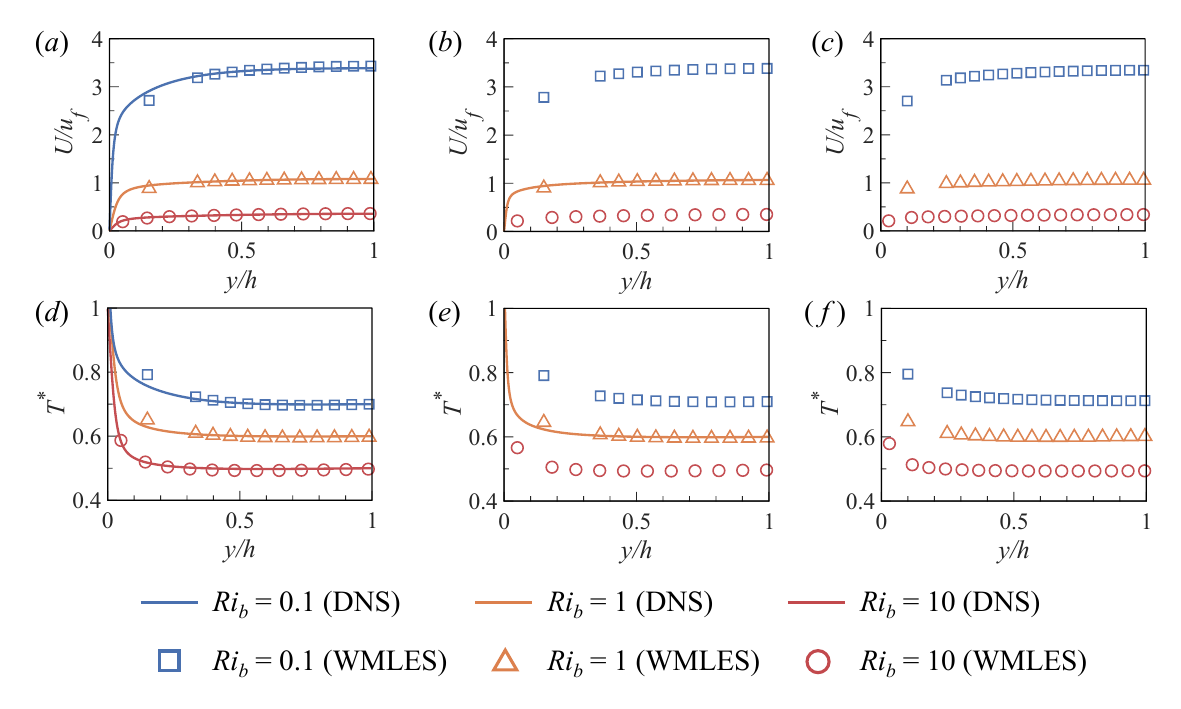}}
	\caption{Profiles of (\emph{a}--\emph{c}) mean streamwise velocity $U/u_f$ and (\emph{d}--\emph{f}) dimensionless mean temperature $T^*=(T-T_c)/\Delta T$ for (\emph{a},\emph{d}) $Ra=10^8$, (\emph{b},\emph{e}) $Ra=10^9$, and (\emph{c},\emph{f}) $Ra=10^{10}$.
		DNS reference data from \citet{PirozzoliBernardiniVerziccoOrlandi2017} are available for all three $Ri_b$ values ($0.1$, $1$, and $10$) at $Ra=10^8$ in (\emph{a},\emph{d}), and only for $Ri_b=1$ at $Ra=10^9$ in (\emph{b},\emph{e}).
		The remaining profiles, including all $Ra=10^{10}$ cases in (\emph{c},\emph{f}), are WMLES predictions without corresponding DNS reference data.}
	\label{fig:mean-velocity-temperature-profiles}
\end{figure}

Figure~\ref{fig:mean-velocity-temperature-profiles} compares the WMLES mean velocity and temperature profiles with the DNS data of \citet{PirozzoliBernardiniVerziccoOrlandi2017}.
The available DNS profiles cover the lower half-channel, $0<y/h<1$.
All DNS--WMLES profile comparisons below are restricted to this range, with the DNS profiles linearly interpolated onto the WMLES cell-centre locations.
For each case, the single wall-adjacent WMLES cell-centre value next to the heated lower wall is excluded from the error evaluation.
The plane containing these wall-adjacent cell centres serves as the wall-model matching plane.
For the mean streamwise velocity $U / u_f$, mean temperature $T^*$ and the three root-mean-square (r.m.s.) velocity fluctuations, we use the pointwise absolute relative error $ \epsilon_\phi(y)=|\phi_{\mathrm{WMLES}}(y)-\phi_{\mathrm{DNS}}(y)| / |\phi_{\mathrm{DNS}}(y)| $, where $\phi_{\mathrm{WMLES}}$ and $\phi_{\mathrm{DNS}}$ denote the corresponding profiles evaluated at the matched wall-normal locations.
Across the DNS-referenced cases shown in figure \ref{fig:mean-velocity-temperature-profiles}, the maximum pointwise absolute relative errors beyond the wall-model matching plane are $3.6\%$ for the mean streamwise velocity and $1.9\%$ for the mean temperature.
The remaining profiles, including all $Ra=10^{10}$ cases, are predictions of the WMLES framework without corresponding DNS reference data.
Across the WMLES cases at fixed $Ri_b$, the normalised mean-velocity and mean-temperature profiles exhibit only a weak dependence on $Ra$ away from the wall, while more noticeable differences occur in the near-wall region.

\begin{figure}
	\centerline{\includegraphics[width=1\textwidth]{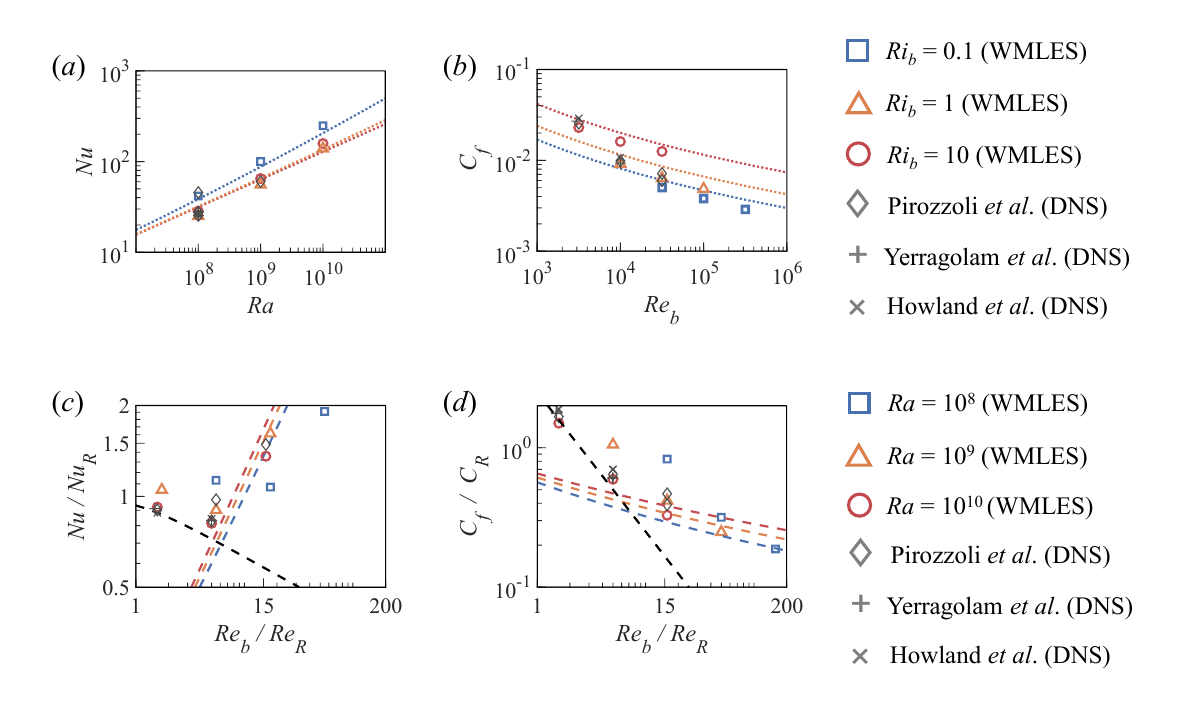}}
	\caption{Variation of (\emph{a}) global Nusselt number $Nu$ with Rayleigh number $Ra$, and (\emph{b}) friction coefficient $C_f$ with bulk Reynolds number $Re_b$.
		(\emph{c}) and (\emph{d}) show the same data as (\emph{a}) and (\emph{b}), respectively, normalised by the corresponding pure Rayleigh--B\'enard reference values $Nu_R$, $C_R$ and $Re_R$ in table \ref{tab:pure-rb-reference-values}.
		Grey symbols show the selected $Pr=1$ DNS data from \citet{PirozzoliBernardiniVerziccoOrlandi2017}, \citet{YerragolamHowlandStevensVerziccoShishkinaLohse2024}, and \citet{HowlandYerragolamVerziccoLohse2024}; the case selection is specified in the text.
		In (\emph{a},\emph{b}), the dotted lines denote the empirical correlations of \citet{PirozzoliBernardiniVerziccoOrlandi2017}.
		In (\emph{c},\emph{d}), the black dashed lines show the buoyancy-dominated predictions of \citet{YerragolamHowlandStevensVerziccoShishkinaLohse2024}, while the coloured dashed lines show the corresponding strong-shear predictions evaluated separately at each $Ra$.}\label{fig:global-transport-scalings}
\end{figure}

We next assess the global heat and momentum transport against the directly matched PRB DNS cases and compare the results with broader external data in figure \ref{fig:global-transport-scalings}.
Across the five directly matched PRB DNS cases listed in table \ref{tab:simulation-parameters}, the maximum relative deviations are $11.7\%$ in $Nu$ for Ra7Re3.5 and $15.9\%$ in $C_f$ for Ra8Re4.5.
For reference, figure \ref{fig:global-transport-scalings}(\emph{a},\emph{b}) also includes the empirical correlations proposed by \citet{PirozzoliBernardiniVerziccoOrlandi2017}, shown as dotted lines.
The correlations for the global Nusselt number and skin-friction coefficient are 
\begin{equation}
	Nu\left(Ra, Re_b\right)=\left[\left(0.1165 Ra^{0.304}\right)^3+\left(0.0073 Re_b^{0.802}\right)^3\right]^{1/3},
\end{equation}
and 
\begin{equation}
	C_f\left(Ra, Re_b\right)=C_f\left(Re_b\right)f\left(Ri_b\right),
\end{equation}
where $C_f\left(Re_b\right)$ follows Prandtl's friction law for turbulent channel flow in the form used by \citet{PirozzoliBernardiniVerziccoOrlandi2017},

\begin{equation}
    \sqrt{\frac{2}{C_f\left(Re_b\right)}}
    =
    \frac{1}{\kappa}
    \ln\left(
        \frac{Re_b}{2}
        \sqrt{\frac{C_f\left(Re_b\right)}{2}}
    \right)
    +C_0-\frac{1}{\kappa},
    \label{eq:prandtl-friction-relation}
\end{equation}
where $\kappa=0.383$ and $C_0=4.17$.
The term $-1/\kappa$ results from converting the local logarithmic velocity profile to the channel bulk velocity.
The buoyancy correction is \citep{PirozzoliBernardiniVerziccoOrlandi2017}

\begin{equation}
	f\left(Ri_b\right)=\left(1+4.64 Ri_b\right)^{0.26}.
\end{equation}
The external DNS data selected for figure \ref{fig:global-transport-scalings} comprise the PRB cases Ra8Re3.5, Ra8Re4, Ra8Re4.5 and Ra9Re4.5 of \citet{PirozzoliBernardiniVerziccoOrlandi2017}; the PRB cases at $Ra=10^8$ with $Re_b=3000$ and $10000$ of \citet{YerragolamHowlandStevensVerziccoShishkinaLohse2024}; and the pressure-driven mixed vertical-convection cases at $Ra=10^8$ with $Re_b=3160$ and $10000$ of \citet{HowlandYerragolamVerziccoLohse2024}.
The imposed-flow Reynolds numbers reported in the external studies are denoted by $Re_b$ here.
The additional data of \citet{YerragolamHowlandStevensVerziccoShishkinaLohse2024} and \citet{HowlandYerragolamVerziccoLohse2024} are included to provide broader transport-trend comparisons, with the vertical-convection data serving as a cross-configuration reference.

Figures \ref{fig:global-transport-scalings}(\emph{c},\emph{d}) show the same data as figures \ref{fig:global-transport-scalings}(\emph{a},\emph{b}), respectively, in the normalised forms $Nu/Nu_R$ and $C_f/C_R$ as functions of $Re_b/Re_R$.
The pure Rayleigh--B\'enard reference values $Nu_R$ and $Re_R$ are taken from the supplementary material of \citet{YerragolamHowlandStevensVerziccoShishkinaLohse2024} at the corresponding $Ra$ and $Pr$.
Here, $Re_R$ is the large-scale-circulation Reynolds number estimated from the r.m.s. velocity in the pure-RB state as $Re_R=U_{\mathrm{rms}}H / \nu$, where $H=2h$ is the full height of the RB cell.
The reference coefficient $C_R$ estimates the response friction coefficient associated with the large-scale-circulation rolls in pure RB convection.
The heat--momentum transport scaling of \citet{YerragolamHowlandStevensVerziccoShishkinaLohse2024} is $Nu\sim Pr^{1/3}C_TRe_T$, where $Re_T$ and $C_T$ characterise the total flow.
For consistency with the normalisation used in figure 3(d) of \citet{YerragolamHowlandStevensVerziccoShishkinaLohse2024}, we take the pure-RB reference coefficient as $C_R=Nu_R / (Re_R Pr^{1/3})$.
The numerical reference values used at $Pr=1$ are listed in table \ref{tab:pure-rb-reference-values}.

\begin{table}
    \begin{center}
        \begin{tabular}{lccc}
            $Ra$  & $10^8$      & $10^9$      & $10^{10}$    \\
            $Nu_R$ & 30.680205   & 61.826991   & 129.776364   \\
            $Re_R$ & 2002.935340 & 5760.012287 & 16489.446203 \\
            $C_R$  & 0.01532     & 0.01073     & 0.00787      \\
        \end{tabular}
        \caption{Pure Rayleigh--B\'enard reference quantities at $Pr=1$ used to normalise the data in figure \ref{fig:global-transport-scalings}.
            The values of $Nu_R$ and $Re_R$ are taken from the supplementary material of \citet{YerragolamHowlandStevensVerziccoShishkinaLohse2024}.}
        \label{tab:pure-rb-reference-values}
    \end{center}
\end{table}

For consistency across configurations, the vertical-convection data of \citet{HowlandYerragolamVerziccoLohse2024} are normalised using the same pure-RB reference quantities $Nu_R$, $Re_R$ and $C_R$ at the corresponding $Ra$, rather than using pure vertical-convection reference values.
This common normalisation is used to facilitate comparison of transport responses across the different mixed-convection configurations.
To interpret these normalised results, we consider the limiting relations proposed by \citet{YerragolamHowlandStevensVerziccoShishkinaLohse2024}.
In their general notation, $Re_S$ and $C_S$ characterise the externally imposed shear; for the present PRB configuration, they correspond to $Re_b$ and $C_f$, respectively.
In the buoyancy-dominated regime, the normalised heat transfer is described by
\begin{equation}
	\frac{Nu}{Nu_R}=\left[
		1+\left(\frac{Re_b}{Re_R}\right)^2
		\right]^{-1/10}.
	\label{eq:yerragolam-buoyancy-heat-transfer}
\end{equation}
Following the reference curve used in figure 3(d) of \citet{YerragolamHowlandStevensVerziccoShishkinaLohse2024}, the buoyancy-dominated friction scaling is represented as
\begin{equation}
	\frac{C_f}{C_R} = 2.5\frac{Re_R}{Re_b}.
	\label{eq:yerragolam-buoyancy-friction}
\end{equation}

In the strong-shear limit, \citet{YerragolamHowlandStevensVerziccoShishkinaLohse2024} express the friction coefficient as

\begin{equation}
    \sqrt{\frac{2}{C_f}}
    =
    \frac{1}{\kappa_Y}
    \ln\left(
        Re_b\sqrt{\frac{C_f}{8}}
    \right)
    +B_Y,
    \label{eq:yerragolam-shear-friction}
\end{equation}
where $\kappa_Y=0.41$ and $B_Y=5$.
Here, $B_Y$ denotes the additive constant in the friction relation and is distinct from the log-law intercept $C_0$ used in equation~(\ref{eq:prandtl-friction-relation}).
The corresponding heat-transfer relation is
\begin{equation}
	Nu = 0.25Pr^{1/2}C_f Re_b.
	\label{eq:yerragolam-shear-heat-transfer}
\end{equation}
The black dashed lines in figures \ref{fig:global-transport-scalings}(\emph{c},\emph{d}) represent the buoyancy-dominated relations (\ref{eq:yerragolam-buoyancy-heat-transfer}) and (\ref{eq:yerragolam-buoyancy-friction}), respectively, whereas the coloured dashed lines represent the corresponding strong-shear relations (\ref{eq:yerragolam-shear-heat-transfer}) and (\ref{eq:yerragolam-shear-friction}), respectively.
The relations are normalised separately using the reference values at each $Ra$.
Although quantitative deviations remain in the transitional region, the two limiting descriptions capture the overall changes in heat transport and wall friction within their respective regimes.

\begin{figure}
	\centerline{\includegraphics[width=1\textwidth]{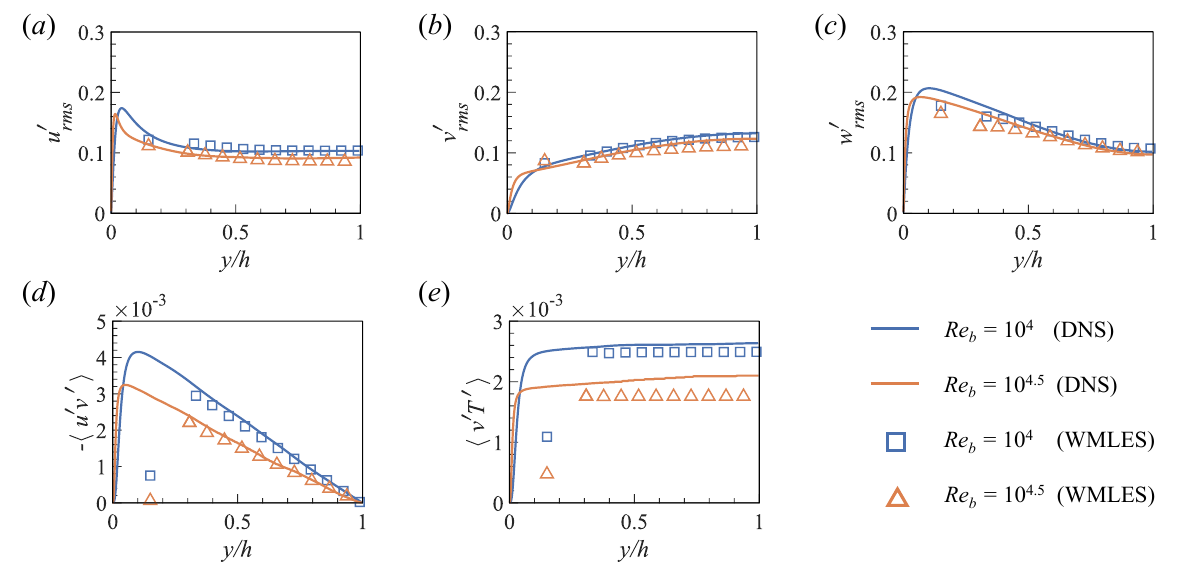}}
	\caption{Wall-normal profiles of the second-order turbulent statistics for $(Ra,Re_b)=(10^8,10^4)$ and $(10^9,10^{4.5})$ at $Ri_b=1$.
		All WMLES statistics are computed from the resolved velocity and temperature fields and are compared with the unfiltered DNS reference statistics of \citet{PirozzoliBernardiniVerziccoOrlandi2017}.
		(\emph{a}--\emph{c}) r.m.s. velocity fluctuations $u^\prime_{\mathrm{rms}}$, $v^\prime_{\mathrm{rms}}$, and $w^\prime_{\mathrm{rms}}$, respectively; (\emph{d}) resolved Reynolds shear stress $-\langle u^\prime v^\prime\rangle_{x,z,t}$; (\emph{e}) resolved wall-normal turbulent heat flux $\langle v^\prime T^\prime\rangle_{x,z,t}$.
		The r.m.s. velocity fluctuations are normalised by $u_f$, the Reynolds shear stress by $u_f^2$, and the turbulent heat flux by $u_f\Delta T$.}
	\label{fig:turbulence-statistics}
\end{figure}

Figure \ref{fig:turbulence-statistics} shows WMLES second-order statistics computed from the resolved velocity and temperature fields.
The DNS profiles of \citet{PirozzoliBernardiniVerziccoOrlandi2017} are used as unfiltered reference statistics.
Figures \ref{fig:turbulence-statistics}(\emph{a}--\emph{c}) compare the r.m.s. velocity fluctuations as functions of wall-normal position at $Ri_b=1$.
Across the DNS-referenced cases shown in figure \ref{fig:turbulence-statistics}, the maximum pointwise absolute relative errors in $u^\prime_{\mathrm{rms}}$, $v^\prime_{\mathrm{rms}}$ and $w^\prime_{\mathrm{rms}}$ are $7.77\%$, $9.86\%$ and $13.86\%$, respectively.
Figures \ref{fig:turbulence-statistics}(\emph{d},\emph{e}) compare the resolved Reynolds shear stress $-\langle u^\prime v^\prime\rangle_{x,z,t}$ and resolved wall-normal turbulent heat flux $\langle v^\prime T^\prime\rangle_{x,z,t}$ from WMLES with the corresponding unfiltered DNS turbulent moments.
Because the DNS Reynolds shear stress approaches zero near the centreline, we assess both transport profiles using the peak-DNS-normalised absolute error, defined as $\epsilon_{\phi,N}(y)= |\phi_{\mathrm{WMLES}}(y)-\phi_{\mathrm{DNS}}(y)| / \max_y|\phi_{\mathrm{DNS}}(y)|$, where $\phi$ denotes either of these transport quantities, and the denominator is the peak magnitude of the corresponding DNS reference profile in the lower half-channel.
The same lower-half-channel restriction, linear interpolation and exclusion of each case's wall-adjacent cell-centre value apply to the evaluation of both error profiles.
The maximum peak-DNS-normalised absolute errors across the two DNS-referenced cases are $7.7\%$ for the resolved Reynolds shear stress and $16.4\%$ for the resolved wall-normal turbulent heat flux.
The near-wall peaks in the resolved Reynolds shear stress are under-resolved on the coarse mesh, whereas the resolved wall-normal turbulent heat flux is underpredicted in the bulk region, particularly at $Ra=10^9$.
We also evaluated the modelled SGS contribution to the wall-normal heat flux. 
Away from the wall-adjacent cell, this contribution is less than approximately 1\% of the total turbulent heat flux, and adding it produces no material change in the comparison with DNS. 
The bulk-region discrepancy at $Ra = 10^9$ therefore cannot be attributed to omission of the modelled SGS flux from figure \ref{fig:turbulence-statistics}(\emph{e}), but instead reflects limitations of the present WALE-based LES treatment and resolution for turbulent thermal transport.

\begin{figure}
	\centerline{\includegraphics[width=1\textwidth]{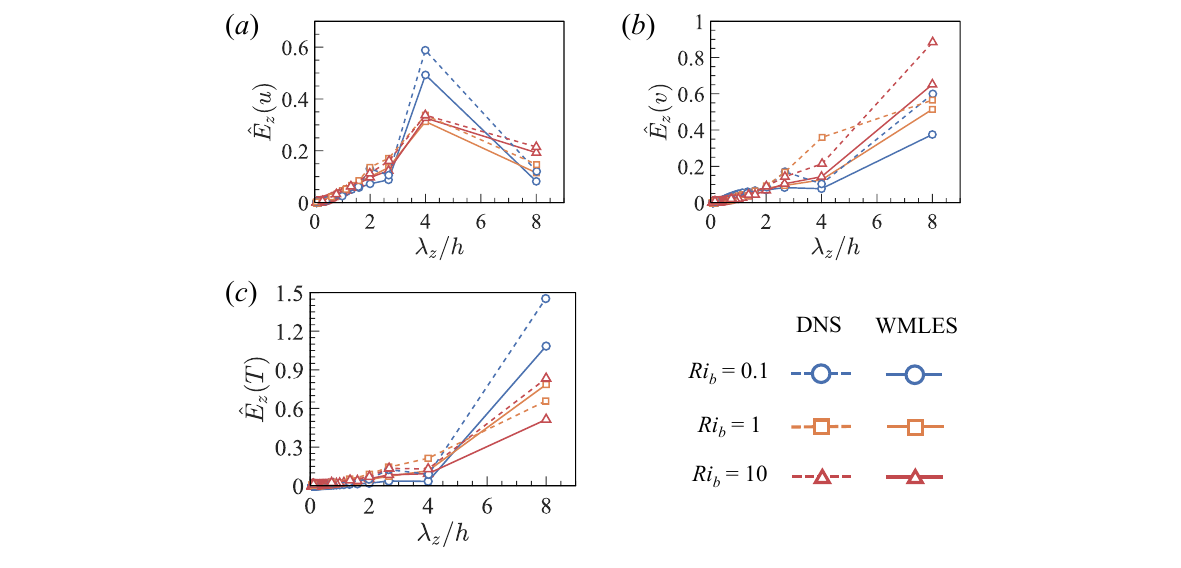}}
	\caption{Nondimensional spanwise spectral densities at the channel centreline.
		Normalised spectra of (\emph{a}) streamwise velocity, (\emph{b}) wall-normal velocity, and (\emph{c}) temperature as functions of spanwise wavelength at $Ra = 10^8$.
		Each spectrum is normalised by the variance of the corresponding quantity.}
	\label{fig:centreline-spanwise-spectra}
\end{figure}

To assess how the WMLES distributes energy across spatial scales, we then compare the spanwise spectra at the channel centreline with the DNS data of \citet{PirozzoliBernardiniVerziccoOrlandi2017}.
As shown in figure \ref{fig:centreline-spanwise-spectra}, the WMLES reproduces the pronounced spectral peak of the streamwise-velocity fluctuations around $\lambda_z/h\approx4$ for all three bulk Richardson numbers, although some differences in peak magnitude remain.
For the wall-normal velocity and temperature, both the DNS and WMLES spectra generally increase towards the longest resolved spanwise wavelengths, indicating appreciable contributions from large spanwise scales.
The WMLES captures these overall spectral trends, but quantitative discrepancies become more evident at the largest resolved wavelength, $\lambda_z/h=8$, particularly for the wall-normal velocity and temperature spectra.
Thus, the dominant spanwise scale of the streamwise-velocity fluctuations is reproduced more closely than the spectral amplitudes of the large-scale wall-normal-velocity and temperature fluctuations.

\subsection{Mesh-count reduction and DNS extrapolation}

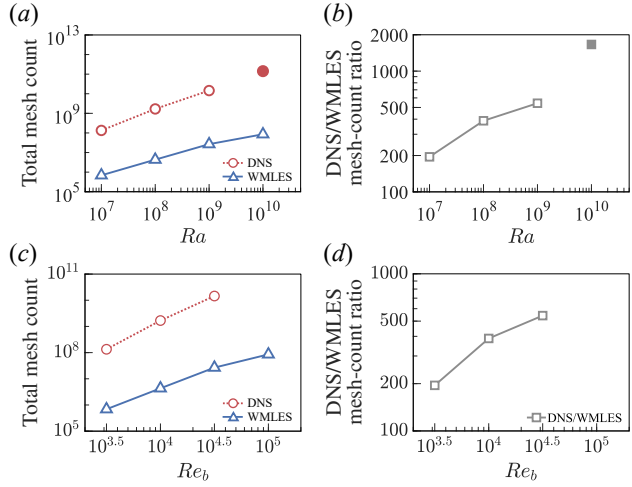
\begin{figure}
    \begingroup
    \setlength{\unitlength}{1bp}%
    \centerline{\resizebox{\textwidth}{!}{%
        \begin{picture}(566.88,283.56)
            \put(0,0){\includegraphics[width=566.88bp]{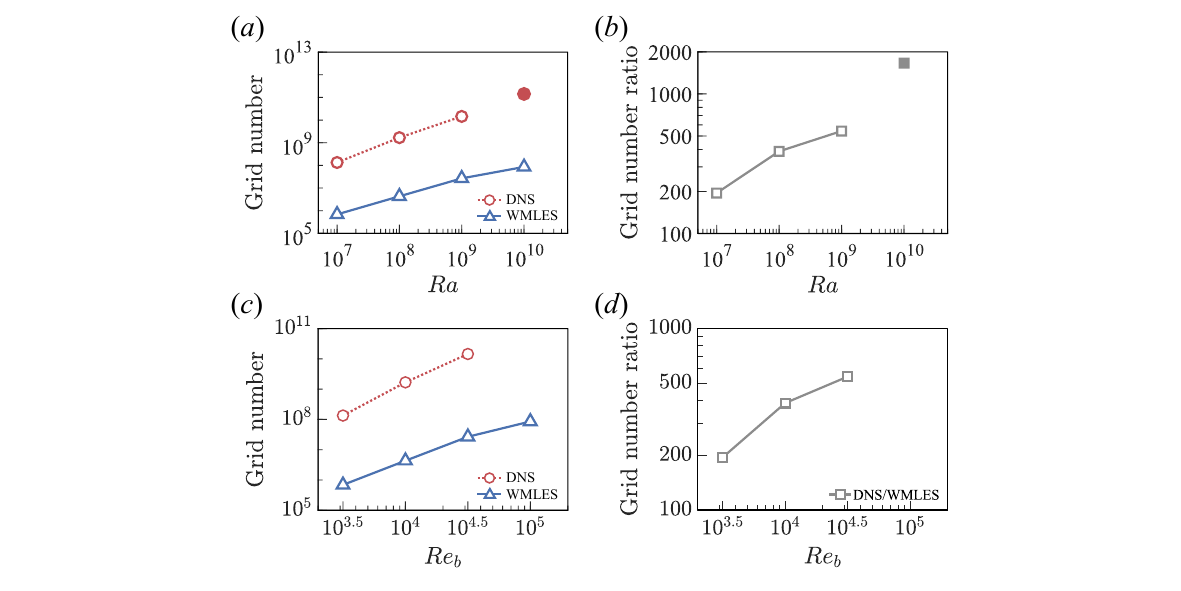}}
            \put(117,182.26){\color{white}\rule{10bp}{65.8bp}}
            \put(297,168.26){\color{white}\rule{10bp}{93.8bp}}
            \put(117.4,49.36){\color{white}\rule{9.7bp}{66bp}}
            \put(297,35.56){\color{white}\rule{10bp}{93.8bp}}
            \fontfamily{ptm}\fontsize{12bp}{12bp}\selectfont
            \color[rgb]{0.149,0.149,0.149}%
            \put(121.8,215.21){\makebox(0,0){\rotatebox[origin=c]{90}{Total mesh count}}}
            \put(296.5,215.22){\makebox(0,0){\rotatebox[origin=c]{90}{\shortstack{DNS/WMLES\\mesh-count ratio}}}}
            \put(122.2,82.40){\makebox(0,0){\rotatebox[origin=c]{90}{Total mesh count}}}
            \put(296.5,82.52){\makebox(0,0){\rotatebox[origin=c]{90}{\shortstack{DNS/WMLES\\mesh-count ratio}}}}
        \end{picture}%
    }}%
    \endgroup
	\caption{Comparison of DNS and WMLES mesh counts.
		(\emph{a}) Total mesh count versus $Ra$.
		(\emph{b}) DNS-to-WMLES mesh-count ratio versus $Ra$.
		(\emph{c}) Total mesh count versus $Re_b$ at $Ri_b=1$.
		(\emph{d}) DNS-to-WMLES mesh-count ratio versus $Re_b$.
		Open circles denote available DNS mesh counts, open triangles the present WMLES mesh counts, and open squares ratios based on available DNS meshes.
		The filled circle in (\emph{a}) and filled square in (\emph{b}) denote the selected extrapolated DNS mesh count and the corresponding ratio for the $Ra=10^{10}$, $Ri_b=0.1$, $Re_\tau\approx6000$ case.
		(\emph{c},\emph{d}) are restricted to the $Ri_b=1$ sequence and do not include this estimate.}
	\label{fig:mesh-count-comparison}
\end{figure}

To quantify the reduction in mesh count, we compare the DNS meshes used by \citet{PirozzoliBernardiniVerziccoOrlandi2017} with the present WMLES meshes in figure \ref{fig:mesh-count-comparison}.
Over $10^7 \le Ra \le 10^9$, the WMLES achieves reductions in mesh count by factors of approximately $195$--$542$ relative to the corresponding DNS meshes.
For the $Ra=10^{10}$ and $Ri_b=0.1$ case ($Re_\tau\approx6000$), no corresponding PRB DNS is presently available.
We therefore estimate a reference DNS mesh count using shear-driven channel-flow and buoyancy-driven convection scalings.

For the shear-driven limit, a fixed domain in outer units and approximately constant streamwise and spanwise resolution in wall units give $N_x,N_z\propto Re_\tau$.
Following the asymptotic wall-normal grid-point scaling discussed by \citet{PirozzoliOrlandi2021}, we adopt $N_y\propto Re_\tau^{3/4}$, giving $N_{\mathrm{ch,ext}}\propto Re_\tau^{2.75}$.
The mesh count of a reference channel DNS in a domain $L_{x,\mathrm{ref}}\times2h\times L_{z,\mathrm{ref}}$ is first rescaled to the present $16h\times2h\times8h$ domain at fixed spatial resolution and then extrapolated in $Re_\tau$, thus
\begin{equation}
        N_{\mathrm{ch,ext}}
        =N_{\mathrm{ref}}
        \frac{16h}{L_{x,\mathrm{ref}}}
        \frac{8h}{L_{z,\mathrm{ref}}}
        \left(\frac{Re_{\tau,\mathrm{target}}}{Re_{\tau,\mathrm{ref}}}\right)^{2.75},
    \label{eq:channel-mesh-extrapolation}
\end{equation}
where $Re_{\tau,\mathrm{target}}=6000$.
The CH4 case of \citet{BernardiniPirozzoliOrlandi2014} has $Re_{\tau,\mathrm{ref}}=4079$, a domain of $6\pi h\times2h\times2\pi h$, and a mesh of $8192\times1024\times4096$.
The LM5200 case of \citet{LeeMoser2015} has $Re_{\tau,\mathrm{ref}}=5186$, a domain of $8\pi h\times2h\times3\pi h$, and a published discretisation count of $10240\times1536\times7680$.
For this spectral DNS, the published streamwise and spanwise counts denote numbers of Fourier modes, and the wall-normal count denotes the number of B-spline basis functions.
Applying equation (\ref{eq:channel-mesh-extrapolation}) to the unrounded mesh products and domain factors gives $N_{\mathrm{ch,ext}}^{\mathrm{CH4}}\approx1.073\times10^{11}$ and $N_{\mathrm{ch,ext}}^{\mathrm{LM5200}}\approx9.747\times10^{10}$; thus, the shear-driven estimate is $N_{\mathrm{ch,ext}}\approx(9.7\text{--}10.7)\times10^{10}$.

For the buoyancy-based extrapolation, the bulk-resolution criterion of \citet{Groetzbach1983} is
\begin{equation}
    \frac{\Delta_{\mathrm{char}}}{H}
    \lesssim\pi\left(\frac{Pr^2}{Ra\,Nu_R}\right)^{1/4},
    \label{eq:rb-bulk-resolution-criterion}
\end{equation}
where $H=2h$ and $\Delta_{\mathrm{char}}$ denotes a characteristic bulk spacing.
At fixed $Pr=1$, this gives $\Delta_{\mathrm{char}}/H\propto[Ra\,Nu_R(Ra)]^{-1/4}$, so the reference Nusselt numbers can determine the change in resolution scale without requiring an additional power-law fit.
We use the pure-RB Nusselt numbers at $Pr=1$ reported by \citet{YerragolamHowlandStevensVerziccoShishkinaLohse2024}, consistent with the reference data in table \ref{tab:pure-rb-reference-values}.
 
We anchor the extrapolation to the $Ra=10^9$ reference grid of \citet{PirozzoliBernardiniVerziccoOrlandi2017}, $N_{\mathrm{ref}}=6144\times768\times3072\approx1.45\times10^{10}$, used for both pure-RB and PRB cases in the same $16h\times2h\times8h$ domain.
Assuming that the domain remains fixed and that all three directions are refined by the same factor implied by the characteristic bulk scale gives
\begin{equation}
    N_{\mathrm{buoy,ext}}
    =N_{\mathrm{ref}}
    \left[
        \frac{Ra_{\mathrm{target}}Nu_R(Ra_{\mathrm{target}})}
             {Ra_{\mathrm{ref}}Nu_R(Ra_{\mathrm{ref}})}
    \right]^{3/4}.
    \label{eq:rb-heat-transport-mesh-extrapolation}
\end{equation}
Using $Ra_{\mathrm{ref}}=10^9$, $Ra_{\mathrm{target}}=10^{10}$ and the corresponding pure-RB Nusselt numbers in table \ref{tab:pure-rb-reference-values} gives $N_{\mathrm{buoy,ext}}\approx1.42\times10^{11}$.
We adopt the larger of the shear-driven and buoyancy-driven estimates as $N_{\mathrm{DNS,ext}} =\max\left(N_{\mathrm{ch,ext}},N_{\mathrm{buoy,ext}}\right) \approx1.42\times10^{11}$.
The present WMLES mesh contains $750\times152\times750=8.55\times10^7$ cells, giving an extrapolated DNS-to-WMLES mesh-count ratio of approximately $1660$.

\subsection{Sensitivity to wall-model parameters}
\begin{figure}
	\centerline{\includegraphics[width=1\textwidth]{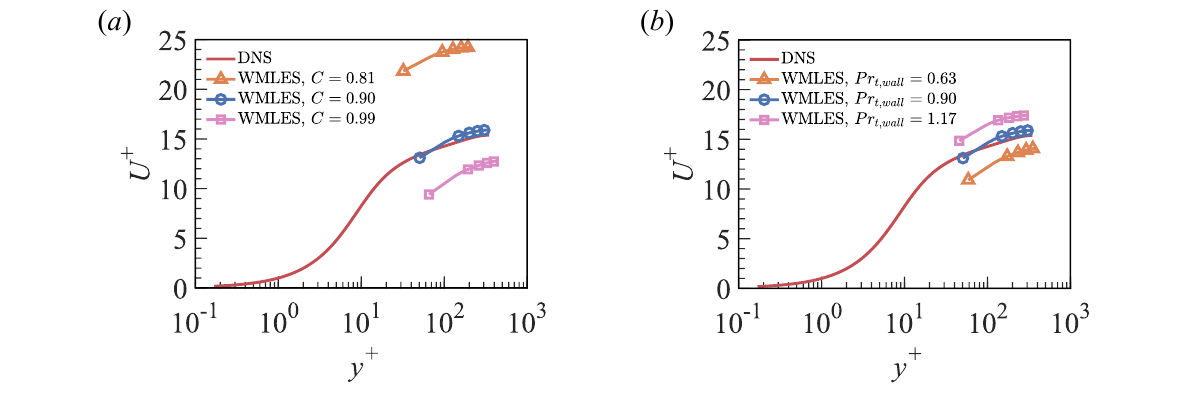}}
	\caption{Sensitivity of the mean streamwise velocity profiles to variations in the wall-model parameters for case Ra8Re4 ($Ra=10^8$, $Re_b=10^4$).
		(\emph{a}) Effect of perturbing the rescaled integration parameter $C$ by $\pm 10\%$ around its baseline value of $0.9$.
		(\emph{b}) Effect of perturbing the wall turbulent Prandtl number $Pr_{t,\mathrm{wall}}$ by $\pm 30\%$ around its baseline value of $0.9$.
		The solid red lines denote DNS data from \citet{PirozzoliBernardiniVerziccoOrlandi2017}.}
	\label{fig:wall-model-parameter-sensitivity}
\end{figure}

We quantify the sensitivity of the WMLES to the prescribed wall-model parameters using case Ra8Re4 as the baseline.
We perturb $C$ by $\pm10\%$ and $Pr_{t,\mathrm{wall}}$ by $\pm30\%$ about their baseline values of $0.9$.
The mean velocity and global transport are sensitive to $C$, as shown in figure \ref{fig:wall-model-parameter-sensitivity}(\emph{a}) and table \ref{tab:wall-model-parameter-sensitivity}.
With $C=0.81$, $Re_\tau$ and $Nu$ are underpredicted relative to DNS by approximately $39\%$ and $61\%$; with $C=0.99$, they are overpredicted by approximately $25\%$ and $67\%$, respectively.
Thus, the calibration of $C$, which affects both the offset and the logarithmic slope of the wall law, has a substantial effect on the predicted transport.
Perturbing $Pr_{t,\mathrm{wall}}$ by $\pm30\%$ produces smaller changes in the mean-velocity profiles than perturbing $C$ by $\pm10\%$ (figure \ref{fig:wall-model-parameter-sensitivity}).
Relative to DNS, the resulting signed relative errors range from $-13.38\%$ to $+11.26\%$ in $Re_\tau$ and from $-21.66\%$ to $+33.67\%$ in $Nu$, as reported in table \ref{tab:wall-model-parameter-sensitivity}.
The calibrated values of $C$ are retained for the investigated $Ri_b$ cases.
The strong transport response to perturbations of $C$ indicates that extrapolation to higher $Ra$ remains sensitive to uncertainty in this parameter.
This sensitivity motivates quantifying the calibration uncertainty in $C$ and developing a predictive closure to replace regime-dependent calibration, potentially incorporating non-equilibrium near-wall effects.
Recent knowledge-integrated approaches combine reduced near-wall physics with data-assisted corrections for effects not represented by equilibrium wall laws \citep{ZhangZhouYangHe2025}, and differentiable WMLES provides a framework for in situ learning and optimisation of wall models using limited reference data \citep{ZhangYangHe2026}.

\begin{table}
	\begin{center}
		\def~{\hphantom{0}}
		\begin{tabular}{lllll}
			Flow case    & $Pr_{t,\mathrm{wall}}$ & $C$  & $Nu$ ($\epsilon_r$) & $Re_\tau$ ($\epsilon_r$) \\
			Ra8Re4-DNS   & -             & -    & 25.443              & 351.01                   \\
			Ra8Re4-WMLES & 0.9           & 0.9  & 25.053 ($-1.53\%$)  & 337.22 ($-3.93\%$)       \\
			Ra8Re4-WMLES & 0.9           & 0.81 & 9.8088 ($-61.45\%$) & 214.12 ($-39.00\%$)      \\
			Ra8Re4-WMLES & 0.9           & 0.99 & 42.523 ($+67.13\%$) & 437.31 ($+24.59\%$)      \\
			Ra8Re4-WMLES & 0.63          & 0.9  & 34.009 ($+33.67\%$) & 390.53 ($+11.26\%$)      \\
			Ra8Re4-WMLES & 1.17          & 0.9  & 19.932 ($-21.66\%$) & 304.05 ($-13.38\%$)      \\
		\end{tabular}
		\caption{Sensitivity of the predicted global flow statistics $Nu$ and $Re_\tau$ to variations in the wall-model parameters.
			The baseline case is Ra8Re4 ($Ra=10^8$, $Re_b=10^4$), for which the baseline parameters are $C=0.9$ and $Pr_{t,\mathrm{wall}}=0.9$.
			The rescaled integration parameter $C$ is perturbed by $\pm 10\%$, whereas the wall turbulent Prandtl number $Pr_{t,\mathrm{wall}}$ is perturbed by $\pm 30\%$.
			DNS data from \citet{PirozzoliBernardiniVerziccoOrlandi2017} are included for comparison.
			The parenthesised values are signed relative errors $\epsilon_r=[\phi_{\mathrm{WMLES}}-\phi_{\mathrm{DNS}}]/\phi_{\mathrm{DNS}}$, expressed as percentages, for $\phi=Nu$ or $Re_\tau$.}
		\label{tab:wall-model-parameter-sensitivity}
	\end{center}
\end{table}

\section{Modulation of turbulent structures by shear and buoyancy}\label{sec:turbulent-structures}

\subsection{Instantaneous flow organisation across regimes}

\begin{figure}
	\centerline{\includegraphics[width=1\textwidth]{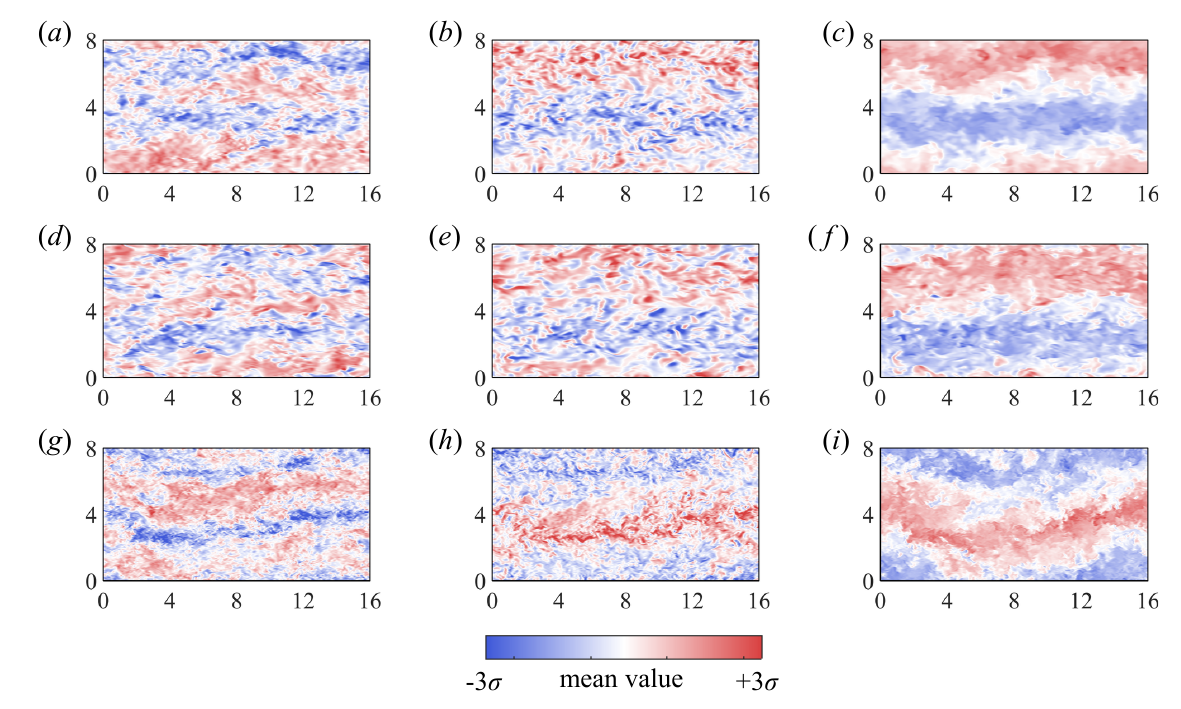}}
	\caption{Instantaneous contours at mid-height for $Ri_b=0.1$: (\emph{a},\emph{d},\emph{g}) streamwise velocity, (\emph{b},\emph{e},\emph{h}) wall-normal velocity, and (\emph{c},\emph{f},\emph{i}) temperature.
		(\emph{a}--\emph{c}) $Ra=10^8$, (\emph{d}--\emph{f}) $Ra=10^9$, (\emph{g}--\emph{i}) $Ra=10^{10}$.
		For each variable, contours are shown within $\pm 3$ standard deviations of the mean.}
	\label{fig:midheight-fields-shear-dominated}
\end{figure}

\begin{figure}
	\centerline{\includegraphics[width=1\textwidth]{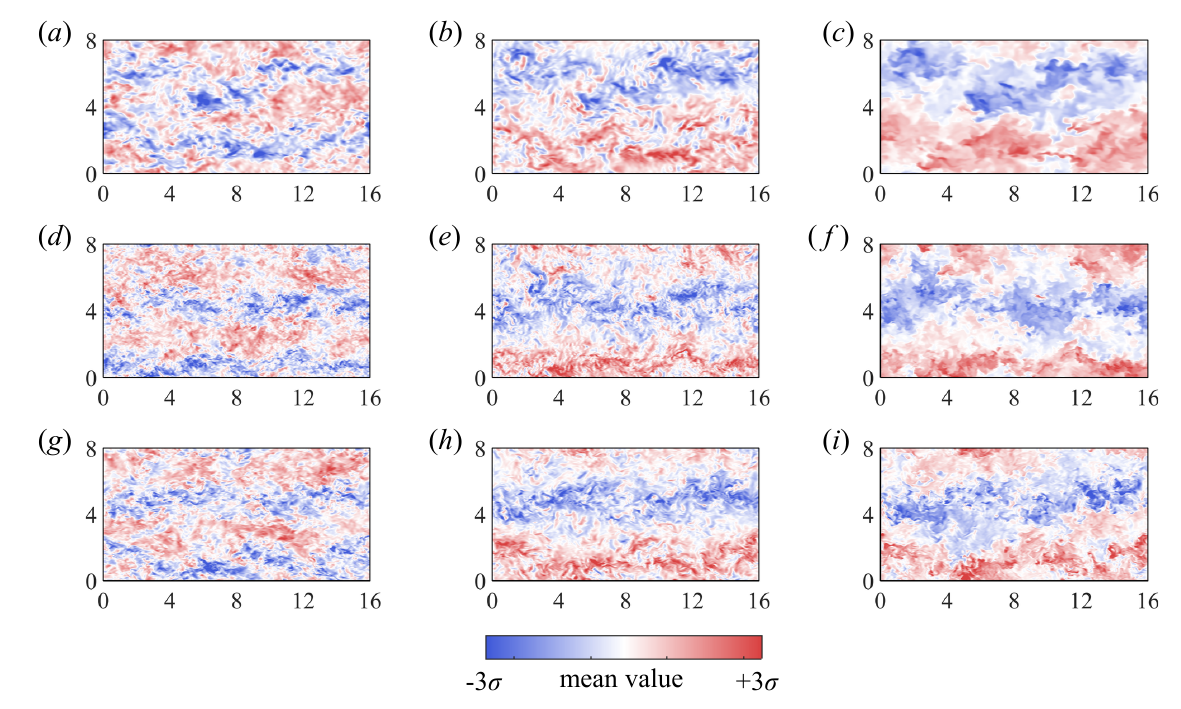}}
	\caption{Instantaneous contours at mid-height for $Ri_b=1$: (\emph{a},\emph{d},\emph{g}) streamwise velocity, (\emph{b},\emph{e},\emph{h}) wall-normal velocity, and (\emph{c},\emph{f},\emph{i}) temperature.
		(\emph{a}--\emph{c}) $Ra=10^8$, (\emph{d}--\emph{f}) $Ra=10^9$, (\emph{g}--\emph{i}) $Ra=10^{10}$.
		For each variable, contours are shown within $\pm 3$ standard deviations of the mean.}
	\label{fig:midheight-fields-comparable-shear-buoyancy}
\end{figure}

\begin{figure}
	\centerline{\includegraphics[width=1\textwidth]{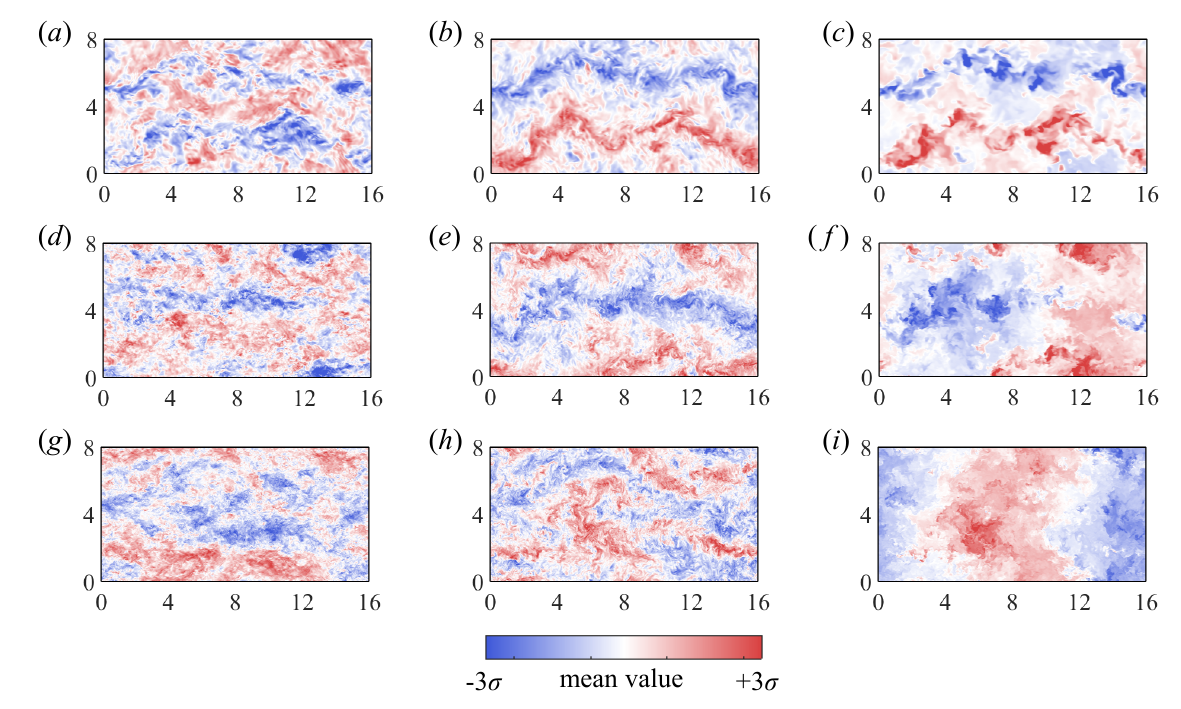}}
	\caption{Instantaneous contours at mid-height for $Ri_b=10$: (\emph{a},\emph{d},\emph{g}) streamwise velocity, (\emph{b},\emph{e},\emph{h}) wall-normal velocity, and (\emph{c},\emph{f},\emph{i}) temperature.
		(\emph{a}--\emph{c}) $Ra=10^8$, (\emph{d}--\emph{f}) $Ra=10^9$, (\emph{g}--\emph{i}) $Ra=10^{10}$.
		For each variable, contours are shown within $\pm 3$ standard deviations of the mean.}
	\label{fig:midheight-fields-buoyancy-dominated}
\end{figure}

We first examine how the instantaneous flow organisation changes with the relative importance of shear and buoyancy.
Figures \ref{fig:midheight-fields-shear-dominated}--\ref{fig:midheight-fields-buoyancy-dominated} show contours of streamwise velocity, wall-normal velocity and temperature at the channel mid-plane ($y/h=1$) for the three $Ri_b$ regimes.
The panels compare cases along each fixed-$Ri_b$ sequence, in which $Ra$, $Re_b$ and $Re_\tau$ increase together while the global ratio of buoyancy to shear remains fixed.
Complementary three-dimensional temporal visualisations are provided in supplementary movies 1--3, showing instantaneous $Q$-criterion isosurfaces coloured by temperature, volume renderings of the streamwise velocity, and volume renderings of the temperature field, respectively. The movies compare the three $Ri_b$ regimes at $Ra = 10^{10}$ and the three Rayleigh numbers along the fixed-$Ri_b = 1$ sequence. 
When shear dominates ($Ri_b=0.1$; figure \ref{fig:midheight-fields-shear-dominated}), the streamwise-velocity and temperature structures remain predominantly elongated in the streamwise direction.
This common alignment is consistent with the influence of streamwise shear on the organisation of both the velocity and temperature fields.
When shear and buoyancy are comparable ($Ri_b=1$; figure \ref{fig:midheight-fields-comparable-shear-buoyancy}), the velocity and temperature streaks remain predominantly streamwise but become less spatially continuous, consistent with the DNS observations of \citet{PirozzoliBernardiniVerziccoOrlandi2017}.
Along the fixed-$Ri_b=10$ sequence (figure~\ref{fig:midheight-fields-buoyancy-dominated}), the streamwise-velocity field changes from fragmented patterns at $Ra=10^8$ and $10^9$ to broader regions at $Ra=10^{10}$, while the wall-normal velocity and temperature fields retain organised spatial features.
We quantify this change below using the streamwise threshold correlation length and the low-wavenumber spectral contribution.


\subsection{Full-field POD and modal energy distribution}

\begin{figure}
	\centerline{\includegraphics[width=1\textwidth]{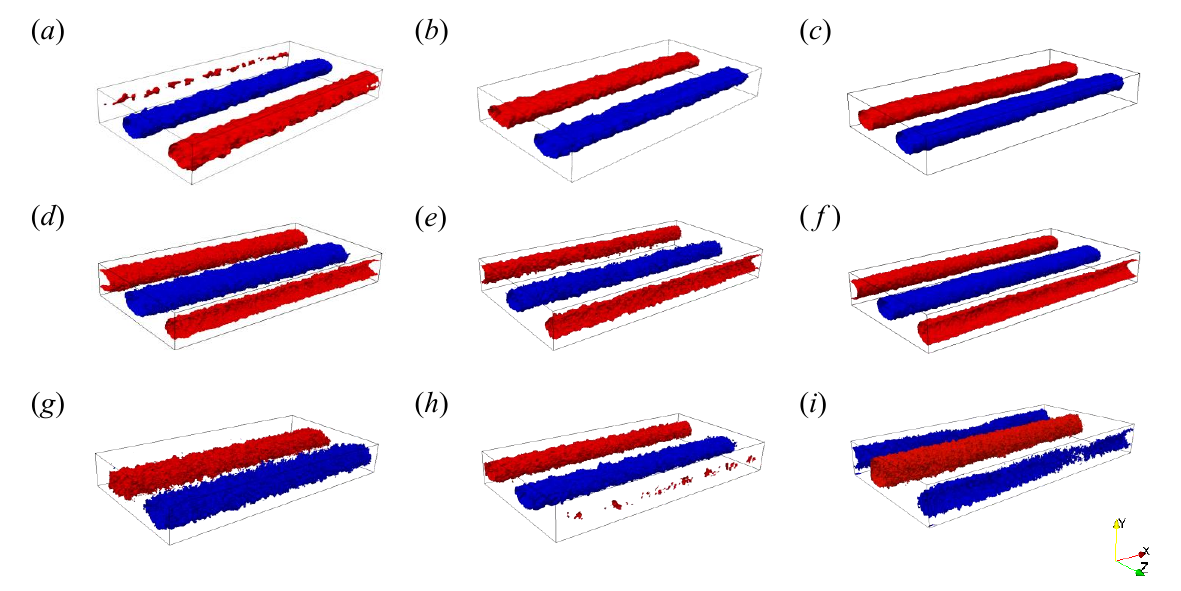}}
	\caption{Isosurfaces of the wall-normal component of the leading full-field POD mode, for (\emph{a}--\emph{c}) $Ra=10^{8}$, (\emph{d}--\emph{f}) $Ra=10^{9}$, and (\emph{g}--\emph{i}) $Ra=10^{10}$, with (\emph{a},\emph{d},\emph{g}) $Ri_b=0.1$, (\emph{b},\emph{e},\emph{h}) $Ri_b=1$, and (\emph{c},\emph{f},\emph{i}) $Ri_b=10$.
		The corresponding friction Reynolds numbers are listed in table~\ref{tab:simulation-parameters}.
		Red and blue denote positive and negative wall-normal components, respectively.}
	\label{fig:leading-pod-mode}
\end{figure}

The instantaneous contours provide a qualitative picture of the large-scale flow organisation.
To identify its dominant energetic components and quantify their relative contributions, we apply full-field proper orthogonal decomposition (POD) to the three-component instantaneous velocity field  $\boldsymbol{u}=(u,v,w)$.
POD has been used to characterise large-scale circulations in convection cells \citep{CastilloCastellanosSergentPodvinRossi2019,SoucassePodvinRiviereSoufiani2019}.
The full velocity field is decomposed into spatial POD modes $\boldsymbol{\varphi}_i(\boldsymbol{x})$ and temporal coefficients $a_i(t)$:

\begin{equation}
	\boldsymbol{u}(\boldsymbol{x}, t) = \sum_{i=1}^{\infty} a_i(t) \boldsymbol{\varphi}_i(\boldsymbol{x}).
\end{equation}
The temporal coefficients satisfy

\begin{equation}
	\langle a_i(t) a_j(t) \rangle_t = \delta_{ij} \lambda_i,
\end{equation}
where $\delta_{ij}$ is the Kronecker delta and $\langle\cdot\rangle_t$ denotes temporal averaging.
The eigenvalues $\lambda_i$ measure the modal contributions to the full-field kinetic energy, and $\lambda_i/\sum_j\lambda_j$ gives the fraction contained in mode $i$.
Figure \ref{fig:leading-pod-mode} shows the wall-normal component of the leading full-field POD mode.
It reveals a streamwise-oriented roll-like organisation in all cases.
This shared modal pattern indicates that coherent roll-like organisation remains identifiable even when the instantaneous streamwise-velocity patterns become more fragmented.

\begin{figure}
	\centerline{\includegraphics[width=1\textwidth]{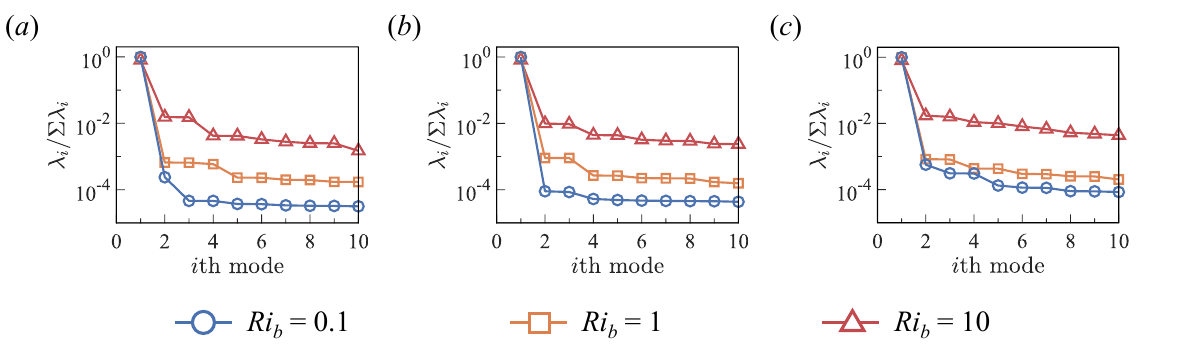}}
	\caption{Full-field kinetic-energy fraction $\lambda_i/\sum_j\lambda_j$ contained in each three-component mode of the full-field POD, for (\emph{a}) $Ra = 10^8$, (\emph{b}) $Ra = 10^9$, and (\emph{c}) $Ra = 10^{10}$.}
	\label{fig:pod-modal-energy-fractions}
\end{figure}

To quantify the modal energy distribution of the full velocity field, we compare the normalised POD eigenvalues $\lambda_i/\sum_j\lambda_j$ across the three flow regimes in figure~\ref{fig:pod-modal-energy-fractions}.
At $Ri_b=0.1$ and $1$, the leading mode accounts for more than $95\%$ of the full-field kinetic energy, largely reflecting the contribution of the mean streamwise flow.
At $Ri_b=10$, this fraction decreases to approximately $80\%$, with a larger fraction distributed among the higher-order modes.
Because the temporal mean is retained in the full-field POD, these modal energy fractions include the mean-flow contribution and should not be interpreted as the kinetic-energy fractions of the roll motions alone.

\begin{figure}
	\centerline{\includegraphics[width=1\textwidth]{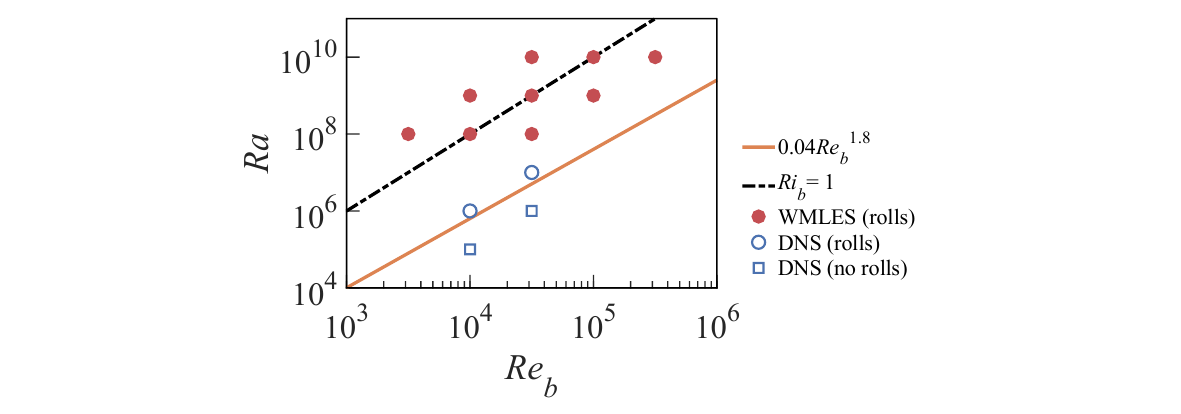}}
	\caption{Rayleigh number $Ra$ versus bulk Reynolds number $Re_b$, showing the predicted critical curve $Ra_{c}(Re_b)$.
		The orange solid line is the fit $Ra_c\approx0.04Re_b^{1.8}$ to the linear-stability predictions of \citet{Cossu2022}.
		Blue open circles and squares denote DNS cases with and without coherent rolls, respectively, from \citet{PirozzoliBernardiniVerziccoOrlandi2017}.
		Red filled circles denote the present WMLES cases, all of which exhibit roll-like structures.
		The black dashed line denotes $Ri_b=1$.}
	\label{fig:critical-rayleigh-number}
\end{figure}

The observed roll-like organisation is consistent with the large-scale rolls discussed in the linear-instability analysis of the turbulent mean flow by \citet{Cossu2022}.
\citet{Cossu2022} argued that the critical Rayleigh number $Ra_c$ for the onset of this linear instability increases with the bulk Reynolds number $Re_b$ according to $Ra_c \approx 0.04Re_b^{1.8}$.
Figure \ref{fig:critical-rayleigh-number} compares this fit with two sampled cases without coherent rolls and the corresponding cases with rolls from \citet{PirozzoliBernardiniVerziccoOrlandi2017}, together with the present WMLES cases.
All present WMLES cases lie above the proposed critical curve and exhibit roll-like structures, consistent with the predicted supercritical organisation.
These structures persist up to $Re_b=10^{5.5}$ in the simulated cases.

\subsection{Correlation lengths and smoothed velocity structures}
\label{sec:threshold-correlation-lengths}

\begin{figure}
	\centerline{\includegraphics[width=1\textwidth]{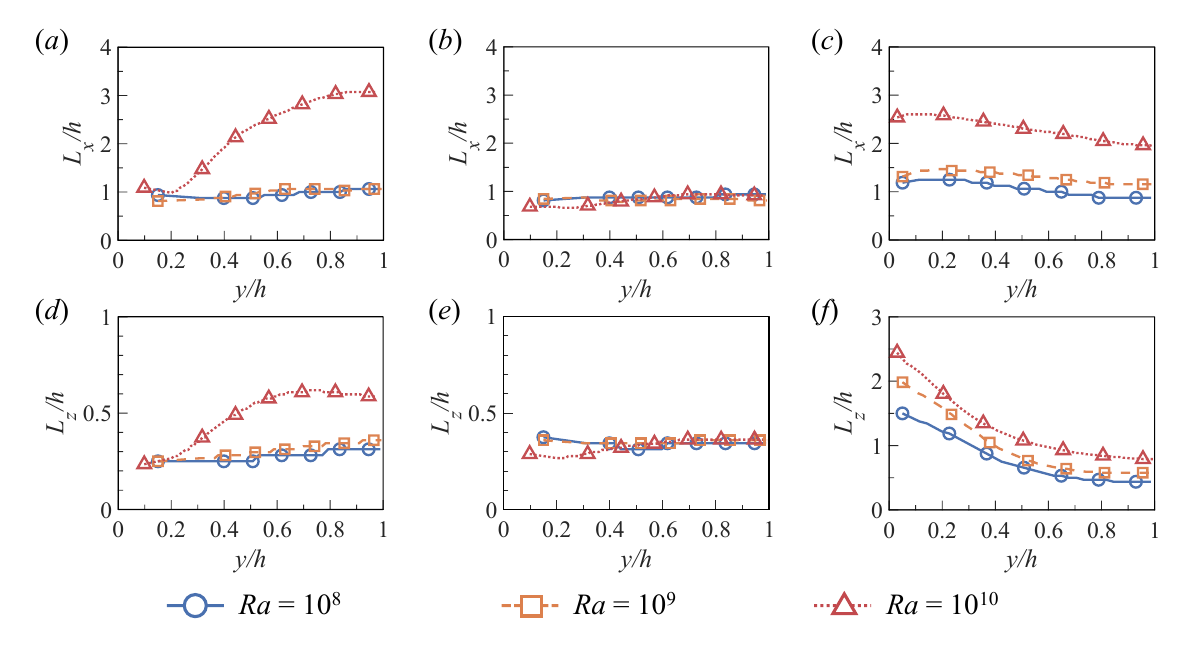}}
	\caption{Wall-normal variations of (\emph{a}--\emph{c}) streamwise threshold correlation length $L_x$ and (\emph{d}--\emph{f}) spanwise threshold correlation length $L_z$.
		(\emph{a},\emph{d}) $Ri_b=0.1$, (\emph{b},\emph{e}) $Ri_b=1$, and (\emph{c},\emph{f}) $Ri_b=10$.
		The lengths are determined from two-point correlations of streamwise velocity fluctuations using a threshold value of $0.15$.}
	\label{fig:threshold-correlation-lengths}
\end{figure}

We next quantify the spatial extent of the flow structures using two-point correlations of the streamwise-velocity fluctuation $u'=u-\langle u\rangle_{x,z,t}$, where $\langle\cdot\rangle_{x,z,t}$ denotes averaging over the homogeneous streamwise and spanwise directions and the sampling interval.
At fixed $y/h$,

\begin{equation}
    R_{uu}(\Delta x,\Delta z;y)
    =
    \frac{\left\langle
    u'(x,y,z,t)\,u'(x+\Delta x,y,z+\Delta z,t)
    \right\rangle_{x,z,t}}
    {\left\langle u'^2(x,y,z,t)\right\rangle_{x,z,t}}.
    \label{eq:streamwise-velocity-correlation}
\end{equation}
The denominator uses the common variance at fixed $y$, since the two points are separated only in statistically homogeneous directions.
One-dimensional correlations are then used to extract threshold correlation lengths.
The streamwise and spanwise correlations are obtained by setting $\Delta z=0$ and $\Delta x=0$, respectively.
The threshold correlation lengths $L_x$ and $L_z$ are defined as the smallest positive separations at which $R_{uu}$ first falls below $0.15$.
These lengths describe the spatial extent of correlated fluctuations.
Figure \ref{fig:threshold-correlation-lengths} shows the wall-normal variations of $L_x$ and $L_z$ for different $Ri_b$.
In the shear-dominated regime ($Ri_b=0.1$, figure \ref{fig:threshold-correlation-lengths}(\emph{a},\emph{d})), $L_x$ is substantially larger than $L_z$, consistent with elongated shear-driven streak-like motions.
Along this fixed-$Ri_b$ path, increasing $Ra$ is accompanied by increasing $Re_b$ and $Re_\tau$, and the streamwise threshold correlation length increases substantially.
At $Ra=10^8$ and $10^9$, the lengths are similar to those at $Ri_b=1$; at $Ra=10^{10}$, $L_x$ reaches approximately $3h$, whereas $L_z$ increases only moderately.
When shear and buoyancy are comparable ($Ri_b=1$, figure \ref{fig:threshold-correlation-lengths}(\emph{b},\emph{e})), both lengths remain relatively small and vary only weakly along the fixed-$Ri_b$ sequence over $Ra=10^8$--$10^{10}$.
Under buoyancy-dominated conditions ($Ri_b=10$, figure \ref{fig:threshold-correlation-lengths}(\emph{c},\emph{f})), both lengths decrease towards the channel centre and vary strongly along the fixed-$Ri_b$ sequence.
At $Ra=10^8$ and $10^9$, $L_x$ remains of order $h$, while $L_z$ decreases from approximately $1.5h$--$2.0h$ near the wall to $0.4h$--$0.6h$ near the centreline.
At $Ra=10^{10}$, $L_x$ reaches approximately $2.0h$--$2.6h$, roughly twice its values at the lower Rayleigh numbers, while $L_z$ decreases from approximately $2.5h$ near the wall to $0.8h$ near the centreline.

\begin{figure}
	\centerline{\includegraphics[width=1\textwidth]{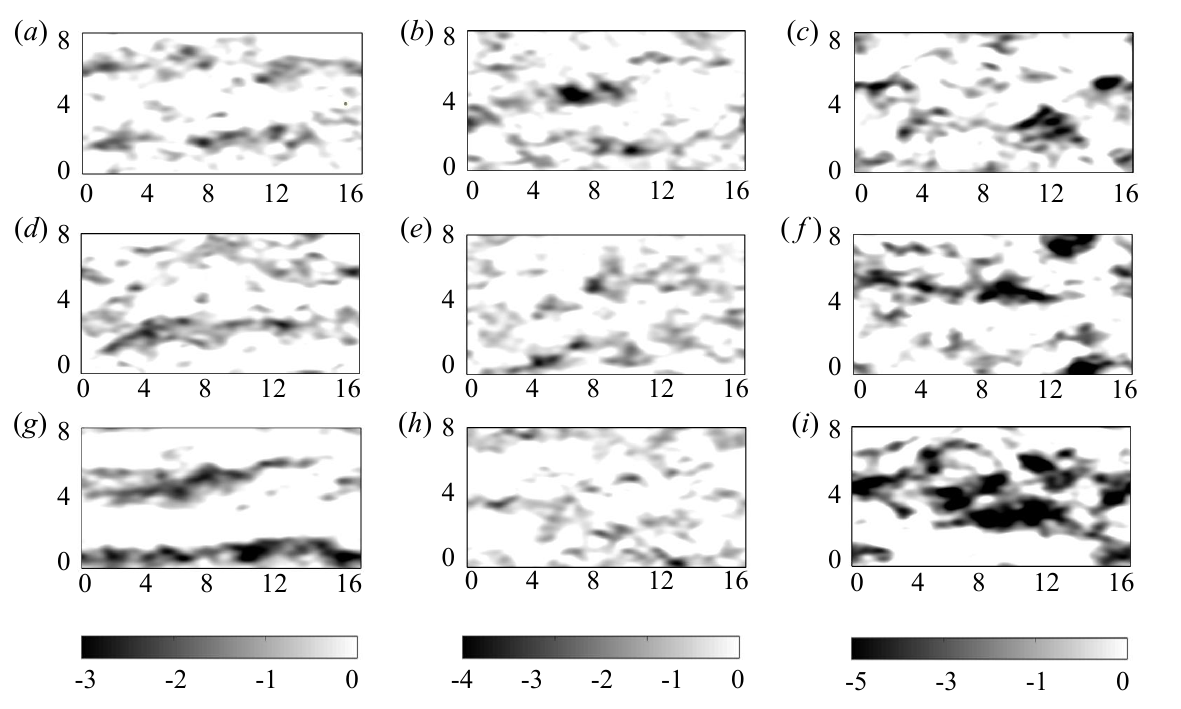}}
	\caption{Spatially smoothed streamwise-velocity fluctuations $u'_G/u_\tau$ at $y/h=1$.
		(\emph{a}--\emph{c}) $Ra=10^8$, (\emph{d}--\emph{f}) $Ra=10^9$, and (\emph{g}--\emph{i}) $Ra=10^{10}$, with (\emph{a},\emph{d},\emph{g}) $Ri_b=0.1$, (\emph{b},\emph{e},\emph{h}) $Ri_b=1$, and (\emph{c},\emph{f},\emph{i}) $Ri_b=10$.
		Within each fixed-$Ri_b$ column, the same colour limits are used for all three Rayleigh numbers; the limits differ among the three $Ri_b$ values as indicated by the colour bars.}
	\label{fig:filtered-streamwise-velocity}
\end{figure}

To visualise the spatial organisation of the low-speed motions, we apply two-dimensional Gaussian smoothing to the streamwise-velocity fluctuation $u'$ at $y/h=1$.
The Gaussian kernel is defined in physical coordinates as
\begin{equation}
    G(x,z)
    =
    \frac{1}{2\pi\sigma_x\sigma_z}
    \exp\left[
        -\frac{x^2}{2\sigma_x^2}
        -\frac{z^2}{2\sigma_z^2}
    \right],
    \label{eq:physical-gaussian-filter}
\end{equation}
with fixed physical standard deviations $\sigma_x=h/8$ and $\sigma_z=h/16$ for all cases.
The Gaussian filter has a gradual spectral response, and the smoothing highlights spatial organisation without defining a sharp separation between large- and small-scale motions.
After truncation at three standard deviations in each direction, the discrete Gaussian kernel is renormalised so that its weights sum to unity.
The convolution is performed periodically in the homogeneous streamwise and spanwise directions.
We denote the smoothed fluctuation by $u'_G$ and plot negative values of $u'_G/u_\tau$ in figure \ref{fig:filtered-streamwise-velocity}.
In the shear-dominated regime ($Ri_b=0.1$), the smoothed low-speed regions become progressively more elongated in the streamwise direction as $Ra$, $Re_b$ and $Re_\tau$ increase along the fixed-$Ri_b$ sequence (figure~\ref{fig:filtered-streamwise-velocity}(\emph{a},\emph{d},\emph{g})).
This trend is consistent with the pronounced increase in $L_x$ shown in figure~\ref{fig:threshold-correlation-lengths}(\emph{a}).
When shear and buoyancy are comparable ($Ri_b=1$), the smoothed fields remain comparatively fragmented over the investigated range (figure~\ref{fig:filtered-streamwise-velocity}(\emph{b},\emph{e},\emph{h})), consistent with the relatively small streamwise and spanwise threshold correlation lengths in figure~\ref{fig:threshold-correlation-lengths}(\emph{b},\emph{e}).
Under buoyancy-dominated conditions ($Ri_b=10$), the smoothed fields exhibit broader low-speed regions whose organisation changes along the fixed-$Ri_b$ sequence (figure~\ref{fig:filtered-streamwise-velocity}(\emph{c},\emph{f},\emph{i})).
The particularly pronounced low-speed regions in the highest-$Ra$ case are consistent with the increased threshold correlation lengths above and the low-wavenumber spectral energy discussed below.

\subsection{Spectral signatures of VLSM-like motions and rolls}

\begin{figure}
	\centerline{\includegraphics[width=1\textwidth]{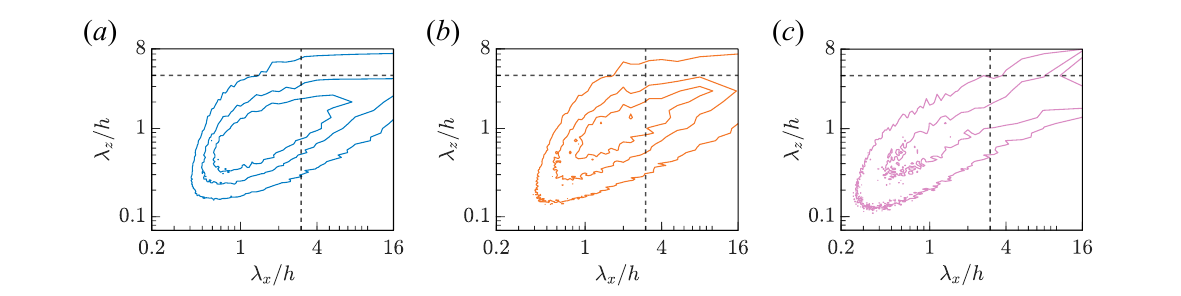}}
	\caption{\hspace{0.25em}Contours of the two-dimensional premultiplied energy spectra of the streamwise velocity fluctuations at $y/h=1$, $k_xk_zE_{xz}(u)/u_\tau^2$, as functions of streamwise wavelength $\lambda_x$ and spanwise wavelength $\lambda_z$.
		(\emph{a}) $Ra = 10^8$, (\emph{b}) $Ra = 10^9$, and (\emph{c}) $Ra = 10^{10}$ for $Ri_b = 0.1$.
		The black horizontal dashed line marks $\lambda_z/h=4$, and the black vertical dashed line marks the reference scale $\lambda_x/h=3$.}
	\label{fig:streamwise-spanwise-premultiplied-spectra}
\end{figure}

To examine whether the $Ri_b=0.1$ cases exhibit signatures consistent with coexisting VLSM-like streamwise-elongated motions and buoyancy-associated rolls, we compute the two-dimensional premultiplied spectra $k_xk_zE_{xz}(u)\left(\lambda_x,\lambda_z\right)$ of the streamwise-velocity fluctuations.
As shown in figure \ref{fig:streamwise-spanwise-premultiplied-spectra}(\emph{a},\emph{b}), at the lower Rayleigh numbers along the fixed-$Ri_b$ sequence, $Ra = 10^8$ and $10^9$, the streamwise fluctuation energy is concentrated mainly at relatively small streamwise and spanwise wavelengths, consistent with the typical footprint of large-scale motions (LSMs).
The reference line at $\lambda_x/h=3$ in figure~\ref{fig:streamwise-spanwise-premultiplied-spectra} highlights the extension of the spectrum towards longer streamwise wavelengths.
At $Ra=10^{10}$, the spectrum extends to longer streamwise and spanwise wavelengths (figure \ref{fig:streamwise-spanwise-premultiplied-spectra}(\emph{c})).
In particular, appreciable spectral energy appears beyond $\lambda_x/h\approx 3$ and around $\lambda_z/h\approx 4$, indicating simultaneous signatures of streamwise elongation and wide spanwise organisation.
Taken together, the large-$\lambda_x$ signature and the spanwise signature near $\lambda_z/h\approx4$ are consistent with the coexistence of VLSM-like streamwise-elongated motions and channel-spanning, buoyancy-associated streamwise rolls at $Ra=10^{10}$.

\begin{figure}
	\centerline{\includegraphics[width=1\textwidth]{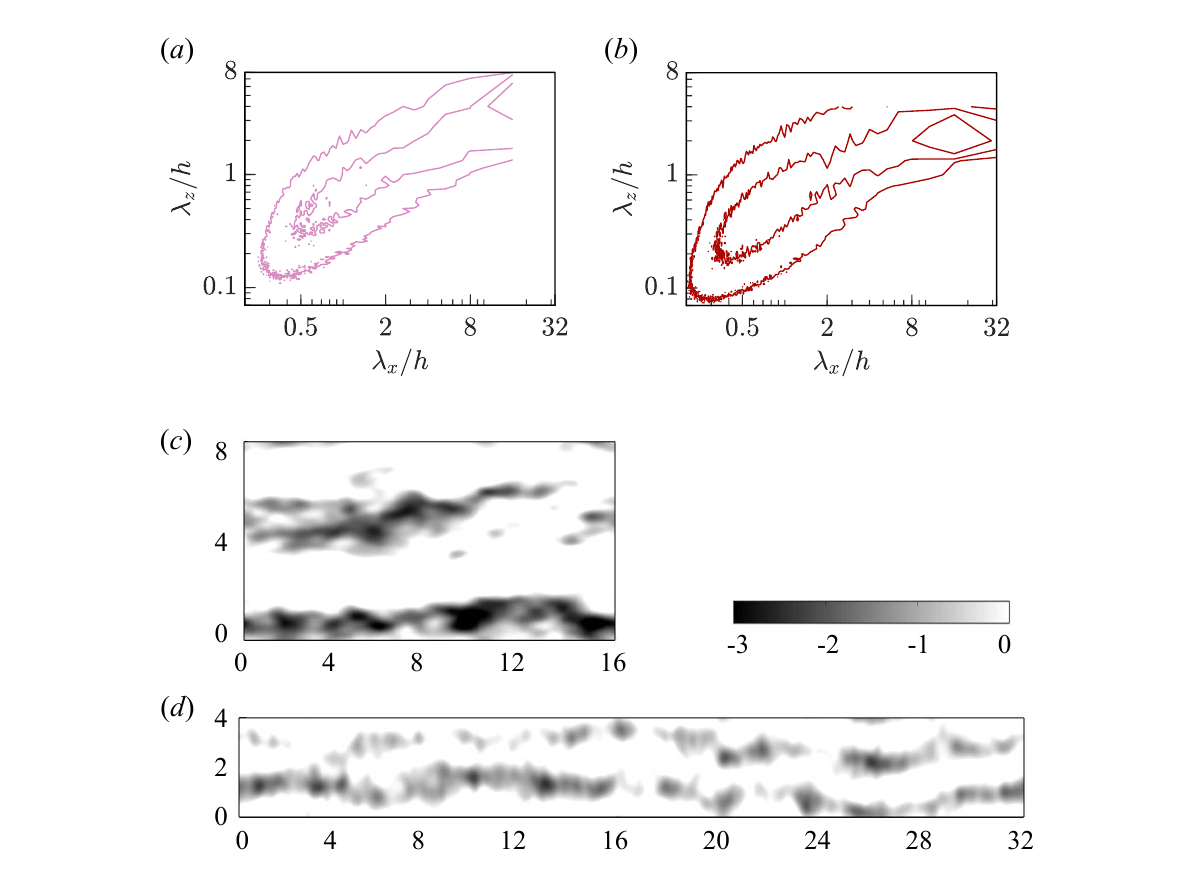}}
	\caption{Comparison of the two-dimensional premultiplied energy spectra and spatially smoothed streamwise-velocity fluctuations between the baseline domain ($16h\times2h\times8h$) and the modified domain ($32h\times2h\times4h$) at $Ra=10^{10}$ and $Ri_b=0.1$.
		(\emph{a},\emph{c}) Baseline domain; (\emph{b},\emph{d}) modified domain.
		(\emph{a},\emph{b}) Two-dimensional premultiplied spectra $k_xk_zE_{xz}(u)/u_\tau^2$ at $y/h=1$.
		(\emph{c},\emph{d}) Spatially smoothed streamwise-velocity fluctuations $u'_G/u_\tau$ at $y/h=1$.
        (\emph{c}) shows the same smoothed snapshot as figure~\ref{fig:filtered-streamwise-velocity}(\emph{g}).
		The same colour limits are used in (\emph{c}) and (\emph{d}), matching figure~\ref{fig:filtered-streamwise-velocity}(\emph{g}).}
	\label{fig:domain-aspect-ratio-comparison}
\end{figure}

At $Ra=10^{10}$ and $Ri_b=0.1$, energetic wavelengths approach the streamwise extent of the baseline domain of $L_{\mathrm{box}}=16h$.
We therefore performed an additional calculation in a modified domain $L\times H\times W=32h\times2h\times4h$, with twice the streamwise length and half the spanwise width of the baseline $16h\times2h\times8h$ domain.
For the visual comparison, we apply the Gaussian smoothing procedure described in \S\ref{sec:threshold-correlation-lengths} to the fields in both domains, with $\sigma_x=h/8$ and $\sigma_z=h/16$.
The modified domain retains both the long-wavelength spectral contribution and the streamwise-elongated low-speed structures (figure~\ref{fig:domain-aspect-ratio-comparison}(\emph{b},\emph{d})).
This supports the persistence of the long-wavelength and streamwise-elongated signatures when the streamwise domain length is doubled.

\begin{figure}
	\centerline{\includegraphics[width=1\textwidth]{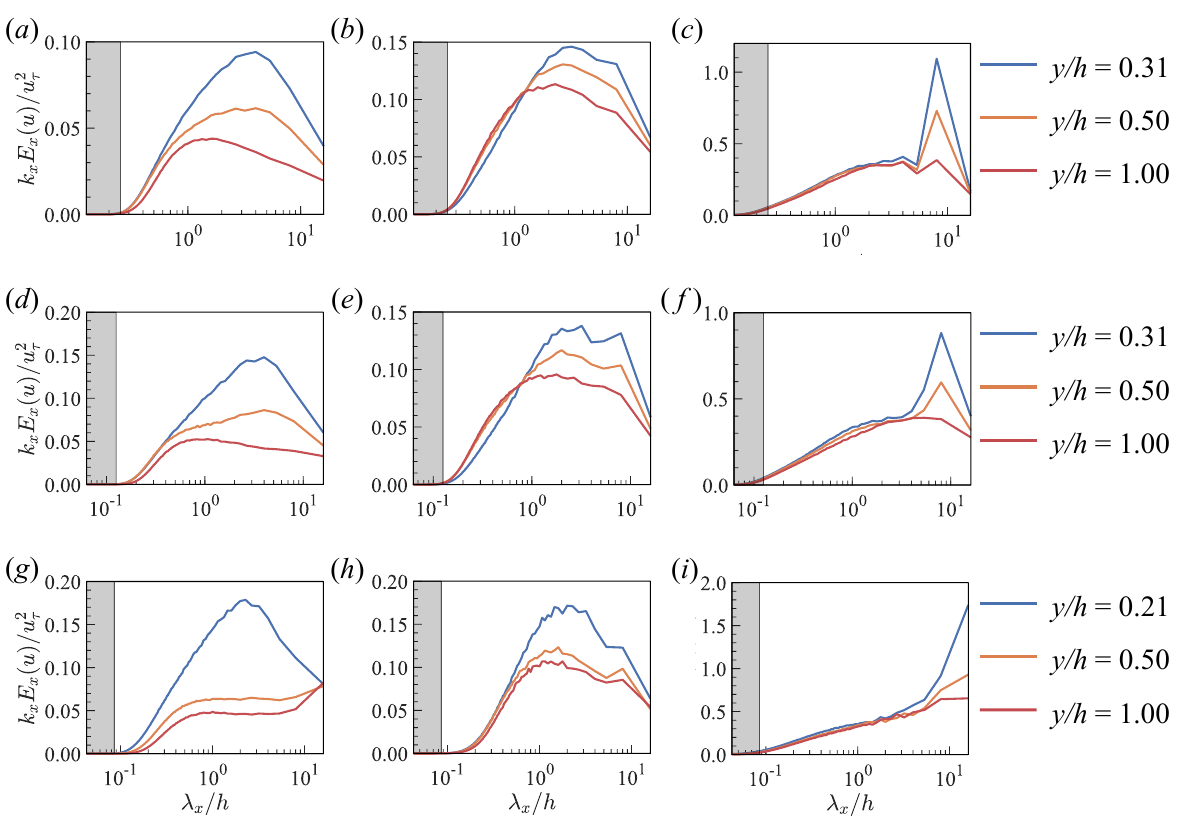}}
	\caption{Premultiplied spectra of the streamwise-velocity fluctuation, $k_xE_x(u)/u_\tau^2$, as functions of streamwise wavelength $\lambda_x/h$.
		(\emph{a}--\emph{c}) $Ra=10^8$, (\emph{d}--\emph{f}) $Ra=10^9$, and (\emph{g}--\emph{i}) $Ra=10^{10}$, with (\emph{a},\emph{d},\emph{g}) $Ri_b=0.1$, (\emph{b},\emph{e},\emph{h}) $Ri_b=1$, and (\emph{c},\emph{f},\emph{i}) $Ri_b=10$.
		The shaded region marks the wavelength range from $\lambda_c$ to $2\lambda_c$, where  $\lambda_c=2\pi/k_c$ is the corresponding cutoff wavelength and $k_c$ is the maximum streamwise wavenumber resolvable by the grid in each LES case.
        Spectral data in this range are excluded from the analysis because numerical errors increase near the grid-resolution limit.}\label{fig:streamwise-spectra-by-height}
\end{figure}

To examine the wall-normal evolution of the spectral contributions, we compare the premultiplied streamwise-velocity spectra at three wall-normal positions in figure \ref{fig:streamwise-spectra-by-height}.
When shear dominates or is comparable to buoyancy ($Ri_b=0.1$ and $Ri_b=1$; left and middle columns of figure \ref{fig:streamwise-spectra-by-height}), the spectra generally exhibit a broad maximum at long wavelengths.
Where a distinct peak is present, it generally shifts towards shorter streamwise wavelengths with increasing wall-normal distance.
At higher $Ra$, $Re_b$ and $Re_\tau$ along the fixed-$Ri_b=0.1$ sequence (figures \ref{fig:streamwise-spectra-by-height}(\emph{d},\emph{g})), the spectral maximum broadens and develops a plateau or shoulder towards larger wavelengths, consistent with the development of VLSM-like long-wavelength contributions.
This trend is qualitatively consistent with the long-wavelength spectral behaviour reported in canonical wall turbulence \citep{KimAdrian1999,DelAlamoJimenezZandonadeMoser2004}.
When buoyancy dominates ($Ri_b=10$; right column of figure \ref{fig:streamwise-spectra-by-height}), the streamwise spectra exhibit a pronounced wall-normal evolution whose detailed form changes along the sequence of increasing $Ra$, $Re_b$ and $Re_\tau$ at fixed $Ri_b$.
At $Ra=10^8$ and $10^9$ (figures \ref{fig:streamwise-spectra-by-height}(\emph{c},\emph{f})), a narrow low-wavenumber peak appears near $\lambda_x/h\approx8$ at the sampling position close to the wall.
Its amplitude decreases at $y/h=0.5$, and the peak becomes substantially weaker and less distinct at the channel centreline.
At $Ra=10^{10}$ (figure \ref{fig:streamwise-spectra-by-height}(\emph{i})), no isolated peak can be identified near $\lambda_x/h\approx8$.
Instead, at the near-wall and intermediate sampling positions, the spectral energy continues to increase towards the largest wavelength permitted by the computational domain, $\lambda_x/h=16$, while the centreline spectrum remains elevated at the longest wavelengths.
Despite these differences in spectral shape, all three $Ri_b=10$ cases show that the low-wavenumber contribution to the streamwise-velocity energy is strongest near the wall and weakens towards the channel centreline.
This wall-normal attenuation is consistent with a weakening streamwise-velocity imprint of buoyancy-associated roll-like organisation.
One interpretation is that wall-normal motions associated with wall-emitted thermal plumes and roll-like organisation redistribute the mean streamwise momentum, generating large-scale streamwise-velocity fluctuations.

\section{Conclusion}\label{sec:conclusion}

In this study, we developed and assessed a buoyancy-modified logarithmic-quadratic wall model for WMLES of incompressible PRB flow at $Pr=1$.
The formulation uses a DNS-motivated, approximately linear near-wall relation between the mean temperature and mean streamwise velocity to construct a thermally modified momentum wall law, together with a thermal wall treatment based on a prescribed wall turbulent Prandtl number.
The temperature--velocity coefficients match the wall value and the ratio of the wall gradients, while DNS comparisons assess the extension of the resulting wall-matched linear relation to finite distances from the wall.
The classical logarithmic law is recovered in the $\beta\rightarrow0$ asymptotic limit with the corresponding limiting behaviour of $C$.
In the \emph{a priori} comparisons, the three wall-law functional forms are calibrated against DNS; the proposed form gives a closer reconstruction of the near-wall velocity and wall-function-equivalent eddy-viscosity profiles than the implemented Businger--Dyer MOST profile and the model of \citet{ScagliariniEinarssonGylfasonToschi2015} under strong buoyancy.
Across the DNS-referenced cases, the maximum pointwise absolute relative errors beyond the wall-model matching plane are $3.6\%$ in the mean velocity and $1.9\%$ in the mean temperature.
Across the cases with available DNS global-transport data, the maximum relative deviations in $Nu$ and $C_f$ are $11.7\%$ and $15.9\%$, respectively.
For cases with corresponding DNS meshes, the reduction factors in mesh count are approximately $195$--$542$.
For the $Ra=10^{10}$ and $Ri_b=0.1$ case ($Re_\tau\approx6000$), for which no corresponding PRB DNS data are available, the selected reference-based extrapolation gives a DNS mesh count of approximately $1.42\times10^{11}$, or 1660 times the present WMLES mesh count.
These reductions in mesh requirements extend the accessible range of flow conditions and enable the modal and spectral analyses of large-scale flow organisation at Reynolds and Rayleigh numbers beyond the presently available PRB DNS database.

The simulations also reveal how shear and buoyancy organise the large-scale flow.
At $Ri_b=0.1$, $1$ and $10$, the wall-normal components of the leading full-field POD modes retain a streamwise-oriented roll-like organisation.
In the full-field POD, stronger buoyancy is accompanied by a larger fraction of the modal energy being distributed among higher-order modes.
For the $Ra=10^{10}$, $Ri_b=0.1$ case, the instantaneous fields, correlation lengths and spectra are consistent with the coexistence of VLSM-like streamwise-elongated motions and buoyancy-associated streamwise rolls.
The spectral analysis further reveals distinct wall-normal trends: the long-wavelength contribution remains appreciable into the outer region in the shear-dominated high-$Re_\tau$ case, whereas under buoyancy-dominated conditions the low-wavenumber streamwise-velocity contribution is strongest near the wall and weakens towards the channel centre.

The present model relies on regime-dependent calibration, with $C=0.90$ for the investigated $Ri_b=0.1$ and $1$ cases and $C=0.95$ for $Ri_b=10$, based on the available PRB DNS database.
The strong sensitivity of the predicted transport to $C$, together with the absence of DNS reference data and of a dedicated grid-refinement study at $Ra=10^{10}$, limits confidence in the quantitative accuracy of the highest-$Ra$ predictions.
Extending the model to other Richardson or Prandtl numbers and geometries will require further validation and, ultimately, a predictive closure for $C$ and the corresponding thermal wall treatment.
Recent Prandtl-number-dependent mean-temperature modelling for incompressible wall turbulence \citep{SunFu2026} and physics-informed or differentiable wall-model frameworks \citep{ZhangZhouYangHe2025,ZhangYangHe2026} provide relevant methodological directions.
Their extension to buoyancy-coupled active-scalar PRB flow, however, remains to be established.

\begin{bmhead}[Supplementary movies]
    Supplementary movies 1--3 are available with the online version of the paper.
\end{bmhead}

\begin{bmhead}[Funding.]
	This work was supported by the National Natural Science Foundation of China (NSFC) under grant nos. 12272311, 12388101 and 12125204; the Young Elite Scientists Sponsorship Program by CAST (2023QNRC001); the Fundamental Research Funds for the Central Universities (no. D5000260269); and the 111 project of China (project no. B17037).
    The authors acknowledge the Computing Center in Xi'an for providing HPC resources that have contributed to the research results reported within this paper.
\end{bmhead}
\begin{bmhead}[Declaration of interests.]
	The authors report no conflict of interest.
\end{bmhead}
\begin{bmhead}[Author ORCIDs.]
	Ao Xu, {\url{https://orcid.org/0000-0003-0648-2701}}; Heng-Dong Xi, {\url{https://orcid.org/0000-0002-2999-2694}}.
\end{bmhead}

\appendix
\begin{appen}

\section{Variations of the pointwise effective parameters}
\label{app:integration-parameter-variations}

To further examine the use of the rescaled integration parameter $C$, we compare the pointwise effective parameters $B_{\mathrm{eff}}(y)$ and $C_{\mathrm{eff}}(y)$ at $Ri_b=1$ for $Ra=10^6$, $10^7$, $10^8$ and $10^9$.
Figures~\ref{fig:integration-parameter-variations}(\emph{a},\emph{b}) show their absolute relative deviations from the corresponding cross-$Ra$ averages, normalised by the magnitudes of those averages.
Here, the overbar denotes an average over the four Rayleigh numbers at each fixed wall-normal position.
This normalisation allows the variations of the two effective parameters to be compared despite their different absolute magnitudes.
For the DNS-referenced WMLES cases at $Ri_b=1$ and $Ra=10^8$ and $10^9$, the wall-model sampling height is $y_p/h=0.15$ (table~\ref{tab:grid-resolution}).
We therefore use $y/h\geq0.15$ as the wall-model-relevant interval when quoting the variations of the effective parameters.
Over this interval, the maximum relative deviation of $B_{\mathrm{eff}}$ is approximately $25\%$, whereas that of $C_{\mathrm{eff}}$ remains below $5\%$.
Figure~\ref{fig:integration-parameter-variations}(\emph{c}) further shows the dependence on $Ra$ of $\beta$ and representative values of the two effective parameters.
For each case, $\langle B_{\mathrm{eff}}\rangle_y$ and $\langle C_{\mathrm{eff}}\rangle_y$ denote arithmetic averages over all sampled wall-normal locations.
As $Ra$ increases, $\beta$ becomes less negative, whereas $\langle B_{\mathrm{eff}}\rangle_y$ becomes more negative.
Since $\beta$ is independent of $y$ within each case and $C_{\mathrm{eff}}=\beta B_{\mathrm{eff}}/2$, these opposing variations partially offset each other, resulting in a substantially weaker variation of $\langle C_{\mathrm{eff}}\rangle_y$ over the available DNS range.

\begin{figure}
	\centerline{\includegraphics[width=1\textwidth]{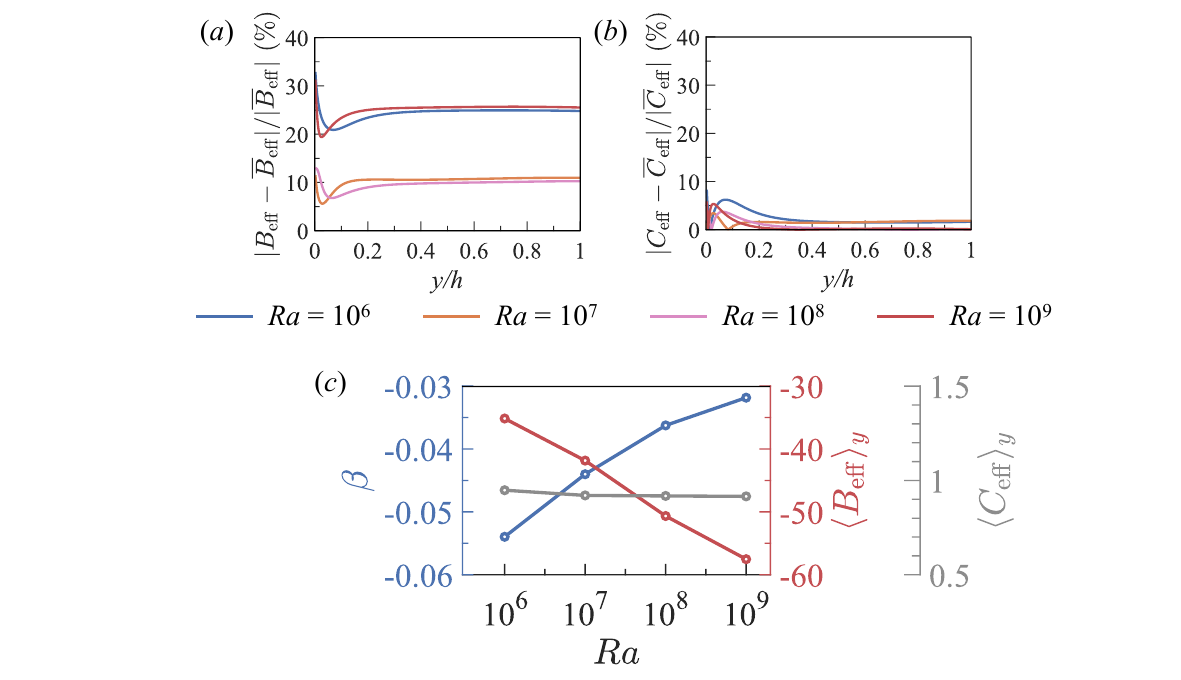}}
	\caption{
	Absolute relative deviations of the pointwise effective parameters (\emph{a}) $B_{\mathrm{eff}}(y)$ and (\emph{b}) $C_{\mathrm{eff}}(y)$ from their corresponding cross-$Ra$ averages at $Ri_b=1$, normalised by the magnitudes of those averages and expressed as percentages.
	The overbar denotes averaging over $Ra=10^6$, $10^7$, $10^8$ and $10^9$ at fixed $y$.
	The wall-model-relevant interval used in the text is $y/h\geq0.15$.
	(\emph{c}) Variations of $\beta$, $\langle B_{\mathrm{eff}}\rangle_y$ and $\langle C_{\mathrm{eff}}\rangle_y$ with $Ra$ at $Ri_b=1$, where $\langle\cdot\rangle_y$ denotes the arithmetic average over all sampled wall-normal locations in $0\leq y/h\leq1$.
	}
	\label{fig:integration-parameter-variations}
\end{figure}

\end{appen}

\bibliographystyle{jfm}

\end{document}